\documentclass[12pt]{article}

\usepackage{header}
\usepackage{abbrev}

\begin{document}

\title{SPHM: a MATLAB package for Smoothed Particle Hydrodynamics simulations}

\author{Marco Sutti\,\orcidlink{0000-0002-8410-1372}\thanks{Mathematics Division, National Center for Theoretical Sciences, Taipei, Taiwan (\email{msutti@ncts.tw}).}}

\date{\today}

\maketitle

\begin{abstract}
We present a MATLAB code that implements the Smoothed Particle Hydrodynamics (SPH) method. 
The paper reviews the continuous Navier--Stokes equations as well as their SPH approximation, adopting a coherent notation that allows to make easy reference to the code. The MATLAB implementation was heavily inspired by the earlier FORTRAN code of G. R. Liu and M. B. Liu, 2003. The code can be used for simple computational fluid dynamics simulations. Two classical benchmark problems are used to validate the algorithm: a one-dimensional shock tube and a two-dimensional shear cavity problem.

\bigskip


\end{abstract}

\tableofcontents

\section{Introduction}
\epigraph{\textit{The particle method is not only an approximation of the continuum fluid equations, but also gives the rigorous equations for a particle system which approximates the molecular system underlying the continuum equations.}}{--- Von Neumann, 1944}

This paper introduces the smoothed particle hydrodynamics (SPH) method, a mesh-free particle method for fluid dynamics simulations.
Its purpose is to serve as a documentation to the SPHM package, which can be downloaded from \url{https://github.com/MarcoSutti/SPHM}. 
SPH has undergone several improvements and has been declined in many variants over the years, and this paper does not mean to be an exhaustive introduction or review to SPH, for which we refer the reader to the extensive review papers~\protect{\cite{Monaghan:2005,Monaghan:2012}}.

This document is designed for scholars who are familiar with the continuous form of the Navier--Stokes equations, and wish to get a quick understanding of how to use the SPH method to obtain their approximated counterparts.


\subsection{Notation}
We list here the notations and symbols adopted, in order of appearance in the paper. Symbols that are only used in one section are typically omitted from the list. Some symbols are inevitably overloaded, but their meaning should be clear from the context.

\begin{table}[htbp]
   \begin{center}
      \begin{tabular}{ll} 
          $ f $                  &  Real-valued function of a real variable\\
          $ \Omega $             &  Domain of SPH simulation\\
          $ \R $                 &  Real numbers \\
          $ d $                  &  Dimension of the problem \\
          $ \bx $, $ \bxprime $  &  Generic particles' positions \\
          $ \delta $             &  Dirac's delta function \\
          $ W $                  &  Smoothing kernel function \\
          $ h $                  &  Smoothing length \\
          $ \nabla $             &  Gradient \\
          $ \nabla \cdot $       &  Divergence \\
          $ a $, $ b $           &  Particle labels \\
          $ \xa $, $ \xb $       &  Positions of particles $ a $ and $ b $, respectively \\
          $ m_{a} $, $ m_{b} $   &  Masses of particles $ a $ and $ b $, respectively \\ 
          $ \rho_{a} $, $ \rho_{b} $ &  Densities of particles $ a $ and $ b $, respectively \\ 
          $ \| \cdot \| $        &  Euclidean norm of a vector \\
          $ W_{ab} $             &  Shortening for the smoothing kernel function $ W(\| \xa - \xb \|, h) $ \\
          $ N $                  &  Number of particles in the support domain of particle $ a $ \\
          $ \nabla_{a} $         &  Gradient with respect to $ \xa $ \\
          $ \rho $               &  Fluid density \\          
          $ \vel $               &  Velocity vector \\ 
          $ V $                  &  Volume of fluid; velocity magnitude\\     
          $ \nabla\vel $         &  Velocity gradient \\       
          $ \bD $                &  Rate of deformation tensor \\
          $ \bW $                &  Vorticity tensor \\
          $ \Sym $               &  Symmetric part of a matrix \\
          $ \tr $                &  Matrix transpose \\
          $ I_{\bD} $            &  First invariant of $ \bD $ \\
          $ \trace $             &  Trace of a matrix \\
          $ \bDprime $           &  Strain deviator \\
          $ \bI $                &  Identity matrix \\
          $ \Skew $              &  Skew-symmetric part of a matrix \\
          $ \bT $                &  Stress tensor \\          
          $ p $                  &  Fluid pressure \\  
          $ \lambda $, $ \mu $   &  Lamé constants \\
          $ K $                  &  Bulk modulus \\
          $ \bTprime $           &  Shear stress tensor \\      
          $ \bb $                &  Body force \\      
          $ \nabla^{2} $         &  Laplacian \\
          $ \Delta p $           &  Pressure difference; spacing between the boundary particles \\
          $ e $                  &  Specific internal energy
      \end{tabular}
   \end{center}
\end{table}

\begin{table}[htbp]
   \begin{center}
      \begin{tabular}{ll} 
          $ T $                  &  Temperature \\
          $ k_{e} $, $ p_{e} $   &  Kinetic and potential energy \\
          $ g $                  &  Gravitational acceleration \\
          $ z $                  &  Elevation \\
          $ _{,j} $              &  Partial derivation with respect to $ x_{j} $ \\
          $ R $                  &  Universal gas constant \\
          $ \gamma $             &  Ratio of specific heats \\
          $ c_{s} $              &  Speed of sound \\
          $ \vba $               &  Relative velocity vector of two particles $a$ and $b$ \\
          $ \bDprimea $          &  Rate of deformation tensor at particle $ a $ \\    
          $ \delta_{ij} $        &  Kronecker delta \\
          $ _{,t} $              &  Partial derivation with respect to time $ t $ \\
          $ \xba $               &  Relative position of two fluid particles $a$ and $b$ \\
          $ h_{0} $              &  Initial smoothing length \\
          $ h_{a} $              &  Smoothing length of particle $ a $ at a certain time instant \\
          $ \Delta t $           &  Time step \\
          $ \kappa $             &  Coefficient determining the size of the support domain\\
          $ \Delta x $           &  Initial spacing between the fluid particles \\
          $ B(a) $               &  Set of boundary particles in the support domain of particle $ a $ \\
          $ \fak $               &  Force per unit mass on fluid particle $ a $ due to boundary particle $ k $ \\
          $ \xak $               &  Relative position of fluid particle $ a $ and boundary particle $ k $ \\
          $ \fk $                &  Force per unit mass on boundary particle $ k $ \\
          $ \fka $               &  Force per unit mass on boundary particle $ k $ due to fluid particle $ a $ \\
          $ \nk $                &  Outward unit normal to a boundary      
      \end{tabular}
   \end{center}
\end{table}

\subsection{Structure of the code and function hierarchy}
\begin{figure}
   \centering
   \includegraphics[width=\textwidth]{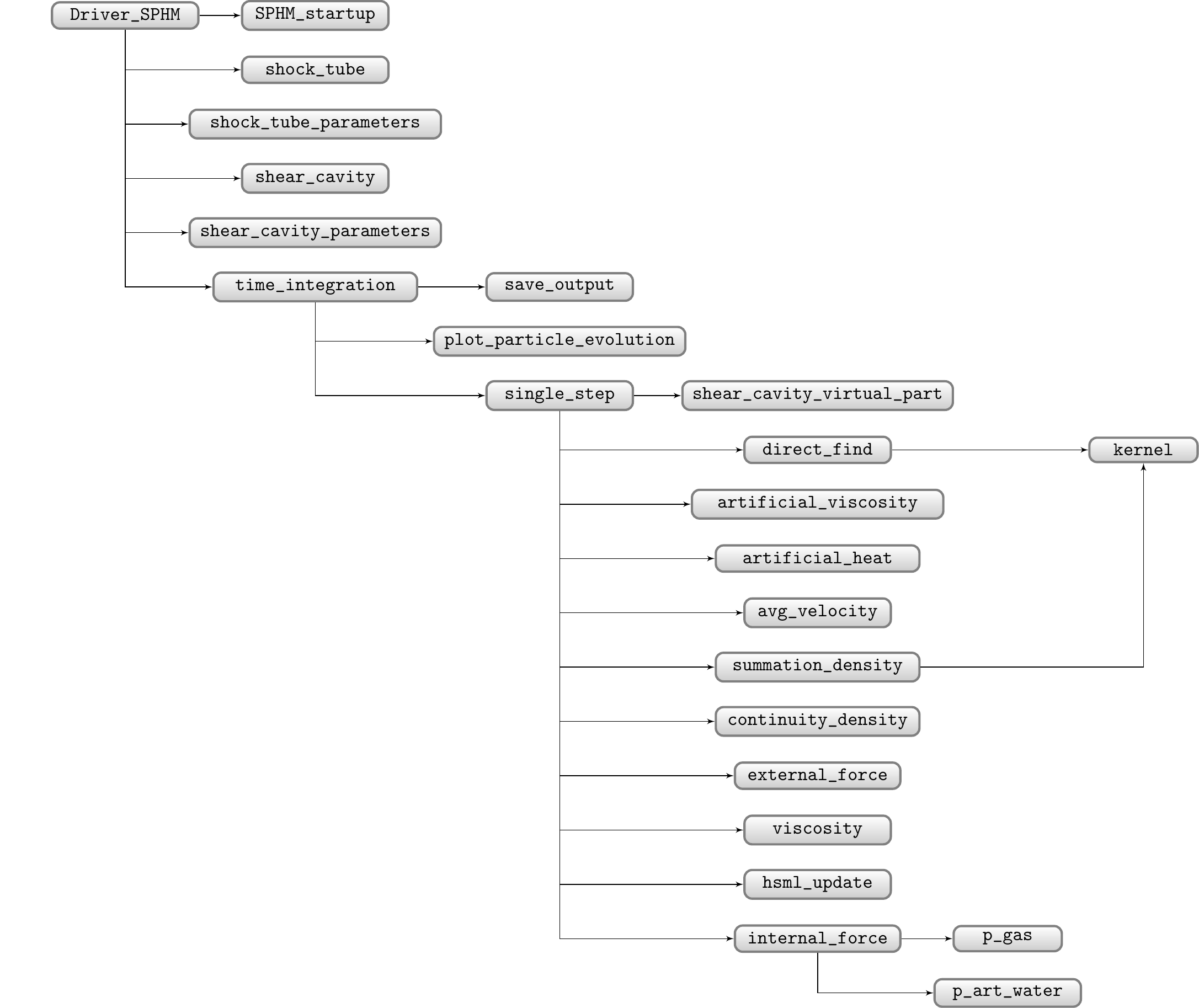}
   \caption{Function hierarchy of SPHM.}
   \label{fig:Function_hierarchy_SPHM}
\end{figure}

Figure~\ref{fig:Function_hierarchy_SPHM} illustrates the function hierarchy of SPHM.

\subsection{Outline}
The remaining part of this document is organized as follows. Section~\ref{sec:SPH_fundamentals} introduces the fundamentals of SPH and of the SPH discretization technique.
Section~\ref{sec:NSE} gives an overview of the continuous Navier--Stokes equations. In section~\ref{sec:SPH_approximation_NSE}, we detail the SPH discretization of Navier--Stokes equations. Section~\ref{sec:art_visc_heat} highlights some other numerical aspects we need to take care of by introducing artificial viscosity and artificial heat.  In section~\ref{sec:boundary_treatment}, we discuss the SPH treatment of boundaries. Numerical experiments are presented in section~\ref{sec:simulations}. In section~\ref{sec:floating_obj} we see how to apply SPH to floating bodies.

\section{SPH fundamentals}\label{sec:SPH_fundamentals}
In this section, we present a short summary of the mathematics behind the SPH discretization technique. We focus on the integral representation of a function, the SPH approximation of the value of a function, the time evolution of the smoothing length, and the time step.

\subsection{Integral representation of a function}
Consider a function $ f \colon \Omega \subset \R^{d} \to \R $, where $d$ is the dimension of the problem, which can be either $1$, $2$, or $3$.
The \emph{integral representation of a function}, which can also be considered as a statement of one of the properties of Dirac's delta function  \protect{\cite[p.~35]{Liu:2003}}, is
\begin{equation}\label{eq:integral_representation}
   f(\bx)=\int_{\Omega}f(\bxprime)\,\delta(\bx-\bxprime)\,\rmd\bxprime.
\end{equation}
A smoothing kernel function is a function $W(\bx-\bxprime,h)$, where $ h $ is called \emph{smoothing length} of the kernel function $ W $, having the special property of mimicking Dirac's delta function when $h$ approaches zero, i.e.,
\begin{equation*}
   \lim_{h \to 0} W(\bx-\bxprime,h)  = \delta(\bx-\bxprime).
\end{equation*}
If we replace Dirac's delta in \eqref{eq:integral_representation} with a smoothing kernel function $W(\bx-\bxprime,h)$ we get
\begin{equation*}
   f(\bx)\doteq\int_{\Omega} f(\bxprime) \, W(\bx-\bxprime,h)\,\rmd\bxprime.
\end{equation*}
The \emph{SPH approximation of the derivative of a function} is
\begin{equation*}
   \langle\nabla\cdot f(\bx)\rangle=\int_{\Omega}\left[ \nabla\cdot f(\bxprime)\right] W(\bx-\bxprime,h)\,\rmd\bxprime,
\end{equation*}
where the symbols $\langle\quad\rangle$ are used to remind us of the approximation that we are committing. In other terms, they are a reminder that the equality holds only if $h \to 0$, i.e., if $ W \to \delta $.

\subsection{SPH approximation of the value of a function}
\begin{figure}[htbp]
\centering
\includegraphics[width=0.70\textwidth]{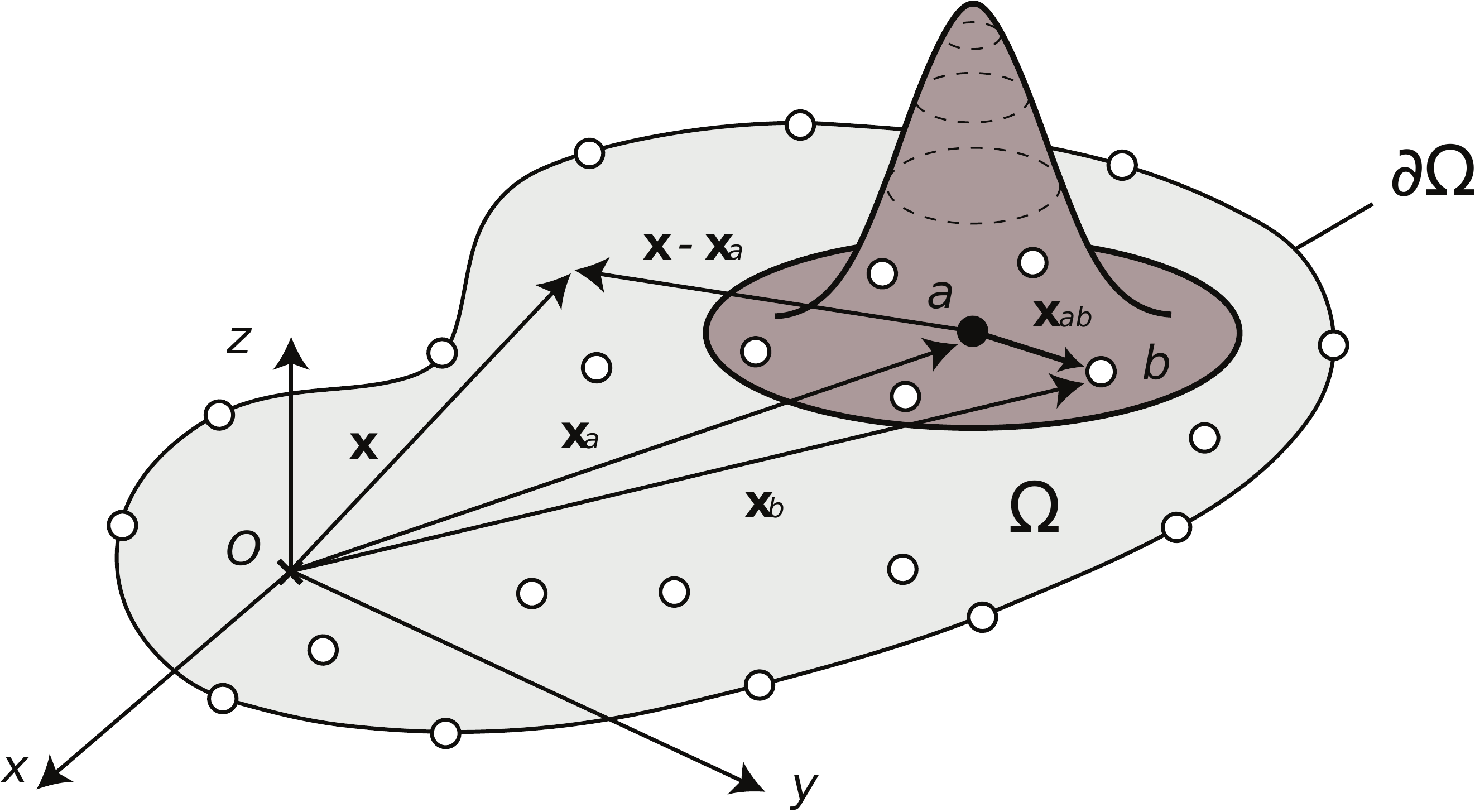} 
\caption{Illustration of SPH configuration on a two-dimensional domain.}
\label{fig:SPH_config}
\end{figure}

Let two SPH particles be labeled $a$ and $b$, and let $\xa, \xb \in \R^{d}$ be their positions. Figure~\ref{fig:SPH_config} depicts a two-dimensional domain $\Omega$ in the $xy$ plane and shows a SPH kernel corresponding to particle $a$. The SPH kernel has a compact support, and a generic particle falling into the support domain of particle $a$ is denoted as $b$.

The \emph{SPH approximation of the value of a function at particle} $a$ is
\begin{equation*}
   \langle f(\xa) \rangle = \sum_{b=1}^{N}\dfrac{m_{b}}{\rho_{b}}\,f(\xb)\,W_{ab}.
\end{equation*}
where $N$ is the number of particles that fall into the support domain of particle $a$, $ m_{b} $, $ \rho_{b} $ are the mass and the density of particle $b$, and $ W_{ab} = W(\| \xa - \xb \|, h) $.
The \emph{SPH approximation of the value of a function derivative at particle} $a$ is \protect{\cite[p.~43]{Liu:2003}}
\begin{equation}\label{eq:SPH_approx_div_vel}
   \langle \nabla\cdot f(\xa) \rangle = \sum_{b=1}^{N}\dfrac{m_{b}}{\rho_{b}}\,f(\xb)\cdot \nabla_{a} W_{ab},
\end{equation}
where $ \nabla_{a} W_{ab} $ indicates the gradient of $ W_{ab} $ with respect to $ \xa $.

\section{Navier--Stokes equations}\label{sec:NSE}
In this section, we review all the components of the continuous Navier--Stokes equations, namely, the continuity equation, the constitutive equations for fluids, the equations of motions and their special forms for particular types of fluids, the energy equation, and the equation of state for gas and artificial water.

The main reference for this section is \protect{\cite{Malvern:1977}}, but \protect{\cite{Romano:2006}} has also been used. For each equation, we typically provide the page or equation number where it can be found in these references. 

\subsection{Continuity equation}
The continuity equation is a statement of the \emph{principle of conservation of mass}. It reads \protect{\cite[eq.~(5.2.7c)]{Malvern:1977}}
\begin{equation}\label{eq:continuity}
   \dfrac{\rmd \rho}{\dt} = -\rho\nabla \cdot \vel,
\end{equation}
where $ \rho $ is the density, $ \vel $ is the velocity, and $\nabla \cdot \vel$ denotes the divergence of the velocity.
Since $\nabla \cdot \vel$ represents the time rate of change of the volume per unit volume, or \emph{volumetric strain}, i.e.,
\begin{equation*}
   \nabla \cdot \vel = \dfrac{1}{\delta V}\dfrac{\D(\delta V)}{\D t},
\end{equation*}
the continuity equation~\eqref{eq:continuity} is telling us that the volume can change only if $\rho$ varies in time, i.e., only if the fluid is compressible. Conversely, if the fluid is incompressible, i.e., $\rho$ is constant, the continuity equation~\eqref{eq:continuity} reduces to
\begin{equation}\label{eq:incompressibility_condition}
   \nabla \cdot \vel =0.
\end{equation}
This version of the continuity equation is called \emph{incompressibility condition}.

\subsection{Constitutive equations for fluids}
In order to describe the kinematics of a fluid flow, we split the \emph{velocity gradient} as \protect{\cite[eq.~(4.4.5b)]{Malvern:1977}}
\begin{equation*}
   \nabla\vel=\bD+\bW,
\end{equation*}
where $\bD$ is the symmetric part of $\nabla\vel$,
\begin{equation}\label{eq:stretching_tensor}
   \bD = \Sym(\nabla\vel) = \tfrac{1}{2}(\nabla\vel + \nabla\vel\tr),
\end{equation}
and it is known as the \emph{rate of deformation} or \emph{stretching tensor}. 
Its definition is due to Euler, 1770.
The so-called \emph{first invariant of} $\bD$ is equal to its trace, which in turn is equal to the divergence of the velocity, i.e.,
\begin{equation*}
   I_{\bD} = \trace(\bD) = \nabla \cdot \vel.
\end{equation*}
Moreover, the \emph{deviator} of $\bD$ is defined as
\begin{equation}\label{eq:devD}
   \bDprime = \bD - \tfrac{1}{3}\,(\trace\bD) \, \bI,
\end{equation}
where $ \bI $ is the identity matrix.
The deviator of the rate of deformation tensor or \emph{strain deviator} measures the change in shape of an element, since it contains the angular deformations. The term $ \frac{1}{3}(\trace\bD)\bI $ is the spherical or \emph{hydrostatic strain}, which represents the volume change.

The \emph{spin} or \emph{vorticity} tensor \textbf{W} is defined as the skew-symmetric part of the velocity gradient
\begin{equation*}
   \bW = \Skew(\nabla\vel) = \tfrac{1}{2}(\nabla\vel - \nabla\vel\tr).
\end{equation*}

In the following, we will review three different formulations of the constitutive equations for fluids.

\subsubsection{Navier--Poisson law of compressible viscous fluids}
The Navier--Poisson law of compressible viscous fluids (also known as Navier--Stokes behavior) relates the stress tensor to the rate of deformation tensor according to the following equation (see \protect{\cite[eq.~(6.3.10)]{Malvern:1977}} or \protect{\cite[eq.~(7.29)]{Romano:2006}})
\begin{equation}\label{eq:compr_visc_fl}
   \bT = -p \bI + \lambda (\trace\bD)\bI + 2\mu \bD,
\end{equation}
where for clarity we omit the dependence on $ \rho $ of $p$, $ \lambda $, and $ \mu $.
The first term in this relationship denotes the static component of the pressure. The second term indicates the dynamic component of the pressure, since it depends on $\trace(\bD)$. The third term represents the shear stress.

The constants $\lambda$ and $\mu$ are two independent parameters characterizing the elastic behavior of a given body. They were introduced by Gabriel Lamé (1795--1870) and thus are often referred to as \emph{Lamé constants}. They have to be determined for every material by means of experiments \protect{\cite[p.~59-60]{Muskhelishvili:1975}}.

\subsubsection{Compressible viscous fluids with no bulk viscosity}
The \emph{bulk modulus} (also known as \emph{modulus of compression} or \emph{bulk viscosity}) is defined as a function of the Lamé constants \protect{\cite[eq.~(6.3.15)]{Malvern:1977}}
\begin{equation*}
   K = \lambda+\tfrac{2}{3}\,\mu.
\end{equation*}
The condition of zero bulk viscosity $ K = 0 $, also known as \emph{Stokes condition} \protect{\cite[eq.~(6.3.17b)]{Malvern:1977}}, implies
\begin{equation}\label{eq:stokes_cond}
   \lambda =-\tfrac{2}{3}\,\mu.
\end{equation}
Under this condition, \eqref{eq:compr_visc_fl} becomes
\begin{align*}
\bT&=-p\bI-\tfrac{2}{3}\,\mu(\trace\bD)\bI+2\mu\bD\\
&=-p\bI+2\mu \left[ \bD-\tfrac{1}{3}\,(\trace\bD)\bI\right].
\end{align*}
Finally, recalling the definition~\eqref{eq:devD}, we get \protect{\cite[eq.~(6.3.23)]{Malvern:1977}}
\begin{equation}\label{eq:compr_visc_fl_no_bulk}
   \bT=\underbrace{-p\bI}_{\substack{\text{isotropic}\\ \text{part of \textbf{T}}}}+\underbrace{2\mu\bDprime}_{\substack{\text{deviatoric}\\ \text{part of \textbf{T}}}}.
\end{equation}
This equation governs the behavior of compressible viscous fluids with no bulk viscosity. The first term in the equation, namely $-p\bI$, is the isotropic part of the stress tensor. The second term is the deviatoric part of the stress tensor, $2\mu\bDprime$, which is called \emph{shear stress tensor} and is denoted as $ \bTprime $ \protect{\cite[eq.~(6.3.24)]{Malvern:1977}}. In fact, $ \bTprime $ is clearly related to shear since $ \bDprime $ contains the angular deformations.

\subsubsection{Incompressible viscous fluids}
To obtain the constitutive equation for incompressible viscous fluids it is sufficient to insert in \eqref{eq:compr_visc_fl_no_bulk} the \emph{incompressibility condition} \eqref{eq:incompressibility_condition}. Under such condition, $\bDprime\equiv\bD$ and therefore \eqref{eq:compr_visc_fl_no_bulk} becomes
\begin{equation}\label{eq:incompr_visc_fl}
\bT = -p\bI + 2\mu\bD.
\end{equation}

\subsection{Equations of motion}
The equations of motion for a continuous medium are derived from the \emph{momentum principle} or \emph{Newton's second law}, and they have the general form \protect{\cite[eq.~(5.30)]{Romano:2006}}
\begin{equation}\label{eq:gen_eom}
   \rho\,\dfrac{\rmd\vel}{\dt}=\nabla\cdot\bT+\rho\bb,
\end{equation}
where $ \bb $ is the body force.
These equations are also known as \emph{Cauchy's equations of motion} \protect{\cite{Romano:2006}}.
Clearly, the specific form of the equation of motion~\eqref{eq:gen_eom} depends on the constitutive relationship for the stress tensor $ \bT $, as we will see in the following subsections.

\subsubsection{Compressible fluids with bulk viscosity}
If we specialize \eqref{eq:gen_eom} by using the constitutive equation \eqref{eq:compr_visc_fl}, we get \protect{\cite[eq.~(9.71)]{Romano:2006}}
\begin{equation*}
\rho\,\dfrac{\rmd\vel}{\dt}= - \nabla p + \nabla(\lambda \nabla\cdot\vel)+\nabla\cdot(2\mu\bD)+\rho \bb.
\end{equation*}
Recalling that, by definition~\eqref{eq:stretching_tensor}, $2\bD=\nabla\vel+\nabla\vel\tr$, the previous equation can also be rewritten as
\begin{align*}
\rho\,\dfrac{\rmd\vel}{\dt}  &= - \nabla p + \nabla(\lambda \nabla\cdot\vel)+\nabla\cdot(\mu\nabla\vel+\mu\nabla\vel\tr)+\rho\bb \\
                             &= - \nabla p + \nabla(\lambda \nabla\cdot\vel)+\mu\nabla^{2}\vel+(\mu\nabla\cdot\vel)\nabla+\rho\bb,
\end{align*}
where $ \nabla^{2} \coloneqq \nabla\cdot \nabla $ is the Laplacian.
Finally we get \protect{\cite[eq.~7.1.14b]{Malvern:1977}}
\begin{equation}\label{eq:eom_compr_fluid_bulk}
   \rho\,\dfrac{\rmd\vel}{\dt}= - \nabla p + (\lambda + \mu)\nabla (\nabla\cdot\vel)+\mu\nabla^{2}\vel+\rho\bb.
\end{equation}

\subsubsection{Compressible fluids with no bulk viscosity}
\label{sec:compressible_no_bulk}
To obtain the equations of motion for compressible fluids with no bulk viscosity it is sufficient to insert the Stokes condition \eqref{eq:stokes_cond} into \eqref{eq:eom_compr_fluid_bulk} and get \protect{\cite[eq.~7.1.17b]{Malvern:1977}}
\begin{equation}\label{eq:eom_compr_fluid_no_bulk}
   \rho\,\dfrac{\rmd\vel}{\dt}= - \nabla p + \tfrac{1}{3}\,\mu\nabla (\nabla\cdot\vel)+\mu\nabla^{2}\vel+\rho\bb.
\end{equation}
This equation can also be obtained from~\eqref{eq:gen_eom} by inserting the constitutive equation for a compressible fluid with no bulk viscosity~\eqref{eq:compr_visc_fl_no_bulk}, i.e.,
\begin{equation*}
\rho\,\dfrac{\rmd\vel}{\dt}= \nabla \cdot (-p\bI+2\mu\bDprime)+\rho\bb.
\end{equation*}
This also provides an alternative expression of the equations of motion, namely,
\begin{equation}\label{eq:eom_compr_fluid_no_bulk2}
   \rho\,\dfrac{\rmd\vel}{\dt}= -\nabla p +\nabla\cdot \bTprime+\rho\bb,
\end{equation}
where $\bTprime=2\mu\bDprime$ is the shear stress tensor.

\subsubsection{Incompressible flows}\label{sec:incompr_flow}
It is sufficient to insert the incompressibility condition \eqref{eq:incompressibility_condition} into \eqref{eq:eom_compr_fluid_no_bulk} and get (see \protect{\cite[eq.~(7.1.17a)]{Malvern:1977}} or \protect{\cite[eq.~(9.75)]{Romano:2006}})
\begin{equation}\label{eq:eom_incompr_fluid}
   \rho\,\dfrac{\rmd\vel}{\dt}= - \nabla p +\mu\nabla^{2}\vel+\rho\bb.
\end{equation}

\subsubsection{Incompressible versus compressible flow analysis}\label{sec:incompr_vs_compr}
In this subsection, we are going to emphasize the differences between incompressible versus compressible flow analysis.
Consider a three-dimensional incompressible flow at constant temperature. In such a case, we have to find four unknowns to completely describe the fluid flow, i.e., the pressure $p$ and the three components of the velocity $\vel$. As a consequence, in an incompressible flow analysis we need four equations: the incompressibility condition $\nabla\cdot \vel=0$ and the three scalar equations of motion from the vector equation \eqref{eq:eom_incompr_fluid}. These equations are coupled since the velocity $\vel$ appears in both of them \protect{\cite{Cengel:2010}}. The pressure and the velocity are also coupled since they both appear in equation \eqref{eq:eom_incompr_fluid}. When the velocity is known, we can integrate the pressure gradient appearing in equation \eqref{eq:eom_incompr_fluid} to compute the pressure field up to an arbitrary constant. To determine this arbitrary constant we need to measure the pressure somewhere in the field, i.e., we need a boundary condition on the pressure. Most of the codes in computational fluid dynamics do not calculate pressure by direct integration of the Navier--Stokes equations, but they use a pressure correction algorithm based on a form of \emph{Poisson's equation} for the pressure difference $\Delta p$~\protect{\cite{Cengel:2010}}
\begin{equation}\label{eq:poisson_eq}
   \nabla^{2}(\Delta p) = \mathrm{RHS}(n).
\end{equation}
Note that since pressure appears only as a gradient in the incompressible Navier--Stokes equations~\eqref{eq:eom_incompr_fluid}, the absolute magnitude of the pressure is not important, only pressure differences matter.
This is not true for compressible flow, where $p$ is the thermodynamic pressure rather than the mechanical one.

To find $p$ in compressible flows, we need to introduce an \emph{equation of state}, which relates the pressure, temperature and density of the fluid (see section~\ref{sec:eos}). Since the equation of state introduces \emph{temperature} as an additional unknown, we also need an additional equation, which expresses the \emph{conservation of energy principle} (see section~\ref{sec:energy}). Therefore, in a compressible flow analysis, we have to deal with six equations in six unknowns, i.e., $ \vel $, $ p $, $ e $, and $ T $.
For simple compressible systems, the total energy consists of internal, kinetic and potential energy, expressed on a unit-mass basis, i.e.,
\begin{equation*}
   e = u + k_{e} + p_{e} = u + \dfrac{1}{2}\,V^2 + gz.
\end{equation*}
The internal energy is the sum of all microscopic forms of energy (energy related to the molecular structure of a system and the degree of molecular activity, also referred to as \emph{thermal energy}). The kinetic energy is the energy that a system possesses as a result of its motion. The potential energy is the energy that a system possesses as a result of its elevation in a gravitational field.

\subsection{Energy equation}\label{sec:energy}
The energy equation is the mathematical formulation of the \emph{first law of thermodynamics}, which expresses the \emph{conservation of energy}:
\begin{quote}
\emph{The time rate of change of the energy inside an infinitesimal fluid cell equals the summation of the heat flux into that fluid cell, and the time rate of work done by the body and surface forces acting on that fluid cell.}
\end{quote}
The general form of the energy equation, neglecting heat flux term and the body force, is (\protect{\cite[eq.~(5.4.7)]{Malvern:1977}} or \protect{\cite[eq.~(5.46)]{Romano:2006}})
\begin{equation}\label{eq:gen_energy}
   \rho\,\dfrac{\de}{\dt} = \bT :\nabla\vel.
\end{equation}
Here, $e$ is a specific energy per unit mass [J/kg] and $:$ denotes the scalar product of two tensors. In index notation, equation \eqref{eq:gen_energy} can be written as
\begin{equation*}
   \rho\,e_{,t}=T_{ij}v_{i,j},
\end{equation*}
with $ i, j = 1, \ldots, d $.

If we consider a compressible viscous fluid with no bulk viscosity, we can use the constitutive law~\eqref{eq:compr_visc_fl_no_bulk} to obtain
\begin{equation*}
   \rho\,\dfrac{\de}{\dt}=(-p\bI+2\mu\bDprime):\nabla\vel,
\end{equation*}
then we obtain~\protect{\cite[eq.~(7.1.14c)]{Malvern:1977}}
\begin{equation}\label{eq:en_compr_visc_fl_no_bulk}
   \rho\,\dfrac{\de}{\dt}=-p(\nabla\cdot\vel)+2\mu(\bDprime:\nabla\vel).
\end{equation}
The meaning of this equation is that the time rate of change of the specific internal energy is equal to the work done by the isotropic pressure multiplying the volumetric strain, plus the energy dissipation due to the viscous shear forces.

\subsection{Equation of state}
\label{sec:eos}
An equation of state is any equation that relates pressure, temperature and density of a substance \protect{\cite{Cengel:2010}}.
In basic applications of SPH, we use equations of state for gas and for artificial water.

\subsubsection{Equation of state for gas}
The most famous equation of state is the ideal gas law \protect{\cite[eq.~(5.6.3a)]{Malvern:1977}}
\begin{equation*}
   p=\rho R T,
\end{equation*}
where $p$ is the absolute pressure, $\rho$ is the density, $R$ is the universal gas constant and $T$ is the absolute temperature. In the SPH calculations, we use the so-called \emph{gamma law} \protect{\cite[eq.~(7.4.14)]{Malvern:1977}}
\begin{equation}\label{eq:eos_gas}
   p=(\gamma - 1)\rho e,
\end{equation}
with the \emph{ratio of specific heats} $ \gamma = 1.4 $. As above, $ e $ denotes the specific internal energy. In  SPHM, this equation of state is implemented in the script \verb+p_gas+. 
Note that if the fluid is incompressible, we have no equation of state, since this is replaced by the condition that the density remains constant.

\subsubsection{Equation of state for artificial water}
In SPH, the water is treated as weakly compressible, thus we refer to this kind of fluid as \emph{artificial water}. This allows the use of an equation of state to determine fluid pressure, which is much faster than solving Poisson's equation \eqref{eq:poisson_eq} as discussed in section~\ref{sec:incompr_vs_compr}. Usually, the compressibility is adjusted so that the time step of the simulation is reasonable. To approximate the real fluid as an artificial compressible fluid, a smaller value than the actual sound speed should be used so that the time step is increased to an acceptable value; but it should also be large enough so that the behavior of the artificial compressible fluid is sufficiently close to the reality \protect{\cite{Liu:2003}}.

The equation of state for artificial water most frequently used when the atmospheric pressure is negligible is \protect{\cite[eq.~(10.1)]{Monaghan:2005}}
\begin{equation*}
   p=B\left[ \left( \dfrac{\rho}{\rho_{0}}\right)^{\gamma}-1 \right],
\end{equation*}
where $\rho_{0}$ is the reference density, $\gamma = 7.0$ and
\begin{equation}\label{eq:B}
   B=\rho_{0}\,\dfrac{c_{s}^{2}}{\gamma}
\end{equation}
is a parameter chosen so that the speed of sound $c_{s}$ is large enough to limit the relative density variation $|\delta \rho|/\rho$. This is done to endow the fluid with a slight compressibility in order to permit the use of an equation of state. This formulation goes back as early as~\protect{\cite[p.~44]{Cole:1948}}.

The sound speed $c_{s}$ is determined according to the \emph{Newton--Laplace formula} for the speed of pressure waves
\begin{equation}\label{eq:NewtonLaplace}
   c_{s} = \sqrt{\dfrac{K}{\rho_{0}}},
\end{equation}
where $K$ is the \emph{bulk modulus}, which measures the substance's resistance to uniform compression. For a gas, $ K=\gamma p $, so that~\eqref{eq:NewtonLaplace} becomes
\begin{equation*}
   c_{s}=\sqrt{\dfrac{\gamma p}{\rho_{0}}}.
\end{equation*}
Usually the sound speed $ c_{s} $ in water at 25$^{\circ}$C is about 1497 m/s for freshwater and 1560 m/s for seawater.

The relative density variation is related to the \emph{Mach number}
\begin{equation*}
   \dfrac{|\delta \rho|}{\rho}\approx M^2,
\end{equation*}
where $ M = v/c_{s} $, with $v$ being the maximum fluid speed. Hence we can ensure a slight density variation, say $|\delta \rho|/\rho\approx 0.01$, if $ M<0.1 $, i.e., if $c_{s}\approx 10\,v$ \protect{\cite{Monaghan:2003}}. This means that the sound speed should be about ten times faster than the maximum fluid velocity in order to keep density variations within less than $ 1\% $ \protect{\cite{Gesteira:2010}}.
In fact, if we substitute $c_{s}= 10\,v$ into~\eqref{eq:B}, we get
\begin{equation*}
   B = 100\,\rho_{0}\,\dfrac{v^2}{\gamma},
\end{equation*}
and the relative density variation is $\approx 0.01$.

As can be seen, the above formulas require an estimation of the maximum fluid speed. For instance, when a dam of height $H$ collapses, an adequate value for the maximum water velocity is $v^2=2gH$ \protect{\cite{Monaghan:1994,Monaghan:2003}} or $v^2=gH$ \protect{\cite{Monaghan:1999}}.
The equation of state for artificial water is implemented in \verb+p_art_water+.

\section{SPH approximation of Navier--Stokes equation}\label{sec:SPH_approximation_NSE}
In this section, we will see how to derive the SPH approximation of the Navier--Stokes equations. The boundary treatment that we will see in the next section is strictly related to the equations of motion.
 
\subsection{SPH approximation of density}
The SPH approximation of density can be performed via two different techniques: the summation density approach and the continuity density approach.

\subsubsection{Summation density}
The SPH approximation of density of particle $a$ with the summation density approach is \protect{\cite[p.~114]{Liu:2003}}
\begin{equation}\label{eq:summation_density}
   \rho_{a} = \sum_{b=1}^{N} m_{b} W_{ab},
\end{equation}
where $N$ is the number of particles that fall into the support domain of particle $a$. This is the preferred formulation for general fluid phenomena. It has the advantage of conserving the mass exactly, but it shows edge effects for particles at the interface between two materials and at the boundaries, an issue known as \emph{boundary particle deficiency}. It is implemented in \verb+summation_density+. 

\subsubsection{Continuity density}
The continuity density approach is the SPH approximation of the continuity equation~\eqref{eq:continuity}. Using~\eqref{eq:SPH_approx_div_vel} to approximate the velocity divergence $ \nabla \cdot \vel $ appearing in~\eqref{eq:continuity}, one can obtain
\begin{equation*}
\dfrac{\rmd\rho_{a}}{\rmd t}=-\rho_{a}\sum_{b=1}^{N}\dfrac{m_{b}}{\rho_{b}}\,v^{ba}_{j}\,\dfrac{\partial W_{ab}}{\partial x_{j}^{a}},
\end{equation*}
where the repeated index $j$ implies summation over it, and $ v^{ba}_{j} $ are the components of the relative velocity $ \vba $. The relative velocity vector $ \vba = \bsym{v^{b}} - \bsym{v^{a}} $ has components $ v^{ba}_{i} = v^{b}_{i} - v^{a}_{i} $, and the position vector  $\xa$ of particle $a$ has components $ x_{i}^{a} $, for $ i = 1, \ldots, d $.
Thanks to the introduction of the velocity difference, this formulation takes into account the relative velocities between two particles, so that the density only changes when particles are in relative motion. The continuity density approach is usually preferred when dealing with discontinuous phenomena, such as explosions or wave-breaking. It is implemented in \verb+continuity_density+. 

\subsection{SPH approximation of the equations of motion}
We recall the equations of motion for compressible fluids with no bulk viscosity~\eqref{eq:eom_compr_fluid_no_bulk2}
\begin{equation*}
   \rho\,\dfrac{\rmd\vel}{\dt}= -\nabla p +\nabla\cdot \bTprime+\rho\bb.
\end{equation*}
If we neglect the body forces, divide both sides by $\rho$, and substitute the constitutive relationship $\bTprime = 2 \mu \bDprime$, we get
\begin{equation*}
   \dfrac{\rmd\vel}{\dt}= -\dfrac{\nabla p}{\rho} + \dfrac{1}{\rho}\,\nabla\cdot(2 \mu \bDprime).
\end{equation*}
In index notation,
\begin{equation*}
   \dfrac{\dv_i}{\dt}= -\dfrac{1}{\rho}\,p_{,i} + \dfrac{2}{\rho}\,\mu\,D'_{ij,j},
\end{equation*}
with the indices $i,j=1,2,3$. As usual, a repeated index implies summation over that index, and $_{,j}$ denotes partial derivation with respect to $x_{j}$.

The SPH approximation of this equation is \protect{\cite[eq.~(4.43)]{Liu:2003}}
\begin{equation}\label{eq:SPH_approx_eom}
\dfrac{\dv_{i}^{a}}{\dt}=-\sum_{b=1}^{N}m_{b}\left( \dfrac{p_{a}}{\rho^{2}_{a}}+\dfrac{p_{b}}{\rho^{2}_{b}}\right) \dfrac{\partial W_{ab}}{\partial x_{i}^{a}} + 2 \sum_{b=1}^{N} m_{b} \left( \dfrac{\mu_{a} D'^{a}_{ij}}{\rho^{2}_{a}}+\dfrac{\mu_{b} D'^b_{ij}}{\rho^{2}_{b}}\right) \dfrac{\partial W_{ab}}{\partial x_{j}^{a}}.
\end{equation}
In the function \verb@internal_force@ of SPHM, the term 
\[
   2 \left( \dfrac{\mu_{a} D'^{a}_{ij}}{\rho^{2}_{a}}+\dfrac{\mu_{b} D'^b_{ij}}{\rho^{2}_{b}}\right) \dfrac{\partial W_{ab}}{\partial x_{j}^{a}}
\]
is stored in the variable \verb+eom_forces+.

If we use the second form of symmetrization for the density \protect{\cite[eq.~(4.42)]{Liu:2003}}, the equations of motion become
\begin{equation*}
\dfrac{\dv_{i}^{a}}{\dt}=-\sum_{b=1}^{N}m_{b}\left( \dfrac{p_{a} + p_{b}}{\rho_{a}\rho_{b}}\right) \dfrac{\partial W_{ab}}{\partial x_{i}^{a}} + 2 \sum_{b=1}^{N} m_{b}\, \dfrac{\mu_{a} D'^{a}_{ij}+\mu_{b} D'^b_{ij}}{\rho_{a}\rho_{b}} \,\dfrac{\partial W_{ab}}{\partial x_{j}^{a}}.
\end{equation*}
Here, $v_{i}^{a}$ denotes the particle $a$ velocity component in the $i$th direction.
$D'^{a}_{ij}$ are the coefficients of the deviator of the rate of deformation tensor for particle $a$.

\subsection{SPH approximation of the deviator}
To complete the SPH approximation of the equations of motion, we need to know how to compute the deviator of the rate of deformation tensor at particle $ a $, $ \bDprimea $. We recall that the rate of deformation tensor is defined as in~\eqref{eq:devD}, i.e.,
\begin{equation*}
   \bDprime = \bD - \tfrac{1}{3}\trace(\bD)\bI.
\end{equation*}
Moreover, $ \bD = \tfrac{1}{2}(\nabla\vel+\nabla\vel\tr) $ and $ \trace(\bD) = \nabla \cdot \vel$, therefore
\begin{equation*}
   \bDprime = \tfrac{1}{2}\,(\nabla\vel+\nabla\vel\tr)-\tfrac{1}{3}\,(\nabla \cdot \vel)\,\bI.
\end{equation*}
In index notation, this expression reads
\begin{equation*}
   D'_{ij} = \tfrac{1}{2}\,(v_{i,j}+v_{j,i}) - \tfrac{1}{3}\,(v_{k,k}) \, \delta_{ij},
\end{equation*}
where $ \delta_{ij} $ is Kronecker's delta.

\paragraph{Remark.} When $ d = 3 $, the deviator $ \bDprime $ has nine components, but six of them are equal to each other because of symmetry properties, so in general one has only six independent components. Similarly, when $ d = 2 $, $ \bDprime $ has four components, but two of them are equal to each other because of symmetry properties, so we are left with only three independent components. 

\bigskip

The SPH approximation for $D'^{a}_{ij}$ is thus \protect{\cite[eq.~(4.48)]{Liu:2003}}
\begin{equation}\label{eq:SPHdevD}
D'^{a}_{ij} = \dfrac{1}{2} \sum_{b=1}^{N} \dfrac{m_{b}}{\rho_{b}}\, v^{ba}_{j}\,\dfrac{\partial W_{ab}}{\partial x_{i}^{a}} + \dfrac{1}{2} \sum_{b=1}^{N} \dfrac{m_{b}}{\rho_{b}}\, v^{ba}_{i}\,\dfrac{\partial W_{ab}}{\partial x_{j}^{a}} - \left(\dfrac{1}{3}\sum_{b=1}^{N} \dfrac{m_{b}}{\rho_{b}}\,v^{ba}_{k}\,\dfrac{\partial W_{ab}}{\partial x_{k}^{a}}\right) \delta_{ij}.
\end{equation}
We expand~\eqref{eq:SPHdevD} and get all the six components of $ \bDprimea $. The following expressions are those appearing in the SPHM function \verb+internal_force+.
\begin{equation*}
D'^{a}_{11} = \sum_{b=1}^{N} \dfrac{m_{b}}{\rho_{b}}\, v^{ba}_{1}\,\dfrac{\partial W_{ab}}{\partial x_{1}^{a}}  -\dfrac{1}{3}\sum_{b=1}^{N} \dfrac{m_{b}}{\rho_{b}}\,\left( v^{ba}_{1}\,\dfrac{\partial W_{ab}}{\partial x_{1}^{a}}+v^{ba}_{2}\,\dfrac{\partial W_{ab}}{\partial x_{2}^{a}}+v^{ba}_{3}\,\dfrac{\partial W_{ab}}{\partial x_{3}^{a}} \right),
\end{equation*}
\begin{equation*}
D'^{a}_{22} = \sum_{b=1}^{N} \dfrac{m_{b}}{\rho_{b}}\, v^{ba}_{2}\,\dfrac{\partial W_{ab}}{\partial x_{2}^{a}}  -\dfrac{1}{3}\sum_{b=1}^{N} \dfrac{m_{b}}{\rho_{b}}\,\left( v^{ba}_{1}\,\dfrac{\partial W_{ab}}{\partial x_{1}^{a}}+v^{ba}_{2}\,\dfrac{\partial W_{ab}}{\partial x_{2}^{a}}+v^{ba}_{3}\,\dfrac{\partial W_{ab}}{\partial x_{3}^{a}} \right),
\end{equation*}
\begin{equation*}
D'^{a}_{33} = \sum_{b=1}^{N} \dfrac{m_{b}}{\rho_{b}}\, v^{ba}_{3}\,\dfrac{\partial W_{ab}}{\partial x_{3}^{a}}  -\dfrac{1}{3}\sum_{b=1}^{N} \dfrac{m_{b}}{\rho_{b}}\,\left( v^{ba}_{1}\,\dfrac{\partial W_{ab}}{\partial x_{1}^{a}}+v^{ba}_{2}\,\dfrac{\partial W_{ab}}{\partial x_{2}^{a}}+v^{ba}_{3}\,\dfrac{\partial W_{ab}}{\partial x_{3}^{a}} \right),
\end{equation*}
\begin{equation*}
D'^{a}_{12} = \dfrac{1}{2}\sum_{b=1}^{N} \dfrac{m_{b}}{\rho_{b}}\, v^{ba}_{2}\,\dfrac{\partial W_{ab}}{\partial x_{1}^{a}} + \dfrac{1}{2}\sum_{b=1}^{N} \dfrac{m_{b}}{\rho_{b}}\, v^{ba}_{1}\,\dfrac{\partial W_{ab}}{\partial x_{2}^{a}} ,
\end{equation*}
\begin{equation*}
D'^{a}_{13} = \dfrac{1}{2}\sum_{b=1}^{N} \dfrac{m_{b}}{\rho_{b}}\, v^{ba}_{3}\,\dfrac{\partial W_{ab}}{\partial x_{1}^{a}} + \dfrac{1}{2}\sum_{b=1}^{N} \dfrac{m_{b}}{\rho_{b}}\, v^{ba}_{1}\,\dfrac{\partial W_{ab}}{\partial x_{3}^{a}},
\end{equation*}
\begin{equation*}
D'^{a}_{23} = \dfrac{1}{2}\sum_{b=1}^{N} \dfrac{m_{b}}{\rho_{b}}\, v^{ba}_{3}\,\dfrac{\partial W_{ab}}{\partial x_{2}^{a}} + \dfrac{1}{2}\sum_{b=1}^{N} \dfrac{m_{b}}{\rho_{b}}\, v^{ba}_{2}\,\dfrac{\partial W_{ab}}{\partial x_{3}^{a}}.
\end{equation*}

\subsection{SPH approximation of energy}
We recall the energy equation~\eqref{eq:en_compr_visc_fl_no_bulk} for compressible viscous fluids with no bulk viscosity
\begin{equation*}
   \rho\,\dfrac{\de}{\dt}=-p\nabla\cdot\vel+2\mu\bDprime:\nabla\vel,
\end{equation*}
in index notation:
\begin{equation*}
   e_{,t}=\dfrac{1}{\rho}\left(-p\,v_{k,k}+2\mu D'_{ij}v_{i,j}\right),
\end{equation*}
where $ i, j, k = 1, \ldots, d $.

The SPH approximation of the energy equation for particle $a$ is \protect{\cite[p.~120]{Liu:2003}}
\begin{equation*}
\dfrac{\de_{a}}{\dt} = \dfrac{1}{2}\sum_{b}^{N} m_{b} \left(\dfrac{p_{a}+p_{b}}{\rho_{a}\rho_{b}} \right) v^{ab}_{j} \dfrac{\partial W_{ab}}{\partial x^{a}_{j}}+2 \,\dfrac{\mu_{a}}{\rho_{a}} \left(D'^{a}_{ij}\right)^2,
\end{equation*}
or, with the second form of symmetrization,
\begin{equation*}
\dfrac{\de_{a}}{\dt} =   \dfrac{1}{2}\sum_{b}^{N} m_{b} \left(\dfrac{p_{a}}{\rho_{a}^2}+\dfrac{p_{b}}{\rho_{b}^2} \right) v^{ab}_{j} \dfrac{\partial W_{ab}}{\partial x^{a}_{j}}+2 \,\dfrac{\mu_{a}}{\rho_{a}} \left(D'^{a}_{ij}\right)^2.
\end{equation*}
where we let $ \left(D'^{a}_{ij}\right)^2 $ denote the sum $ \sum_{i,j=1}^{d} D'^{a}_{ij} \, D'^{a}_{ij} $.

We can write the extended version of the first form to obtain the expression that is implemented in SPHM, namely,
\begin{align*}
\dfrac{\de_{a}}{\dt} &= \underbrace{\dfrac{1}{2}\sum_{b}^{N} m_{b} \left(\dfrac{p_{a}+p_{b}}{\rho_{a}\rho_{b}} \right) \underbrace{\left[ v^{ab}_{1} \dfrac{\partial W_{ab}}{\partial x^{a}_{1}}+v^{ab}_{2} \dfrac{\partial W_{ab}}{\partial x^{a}_{2}}+v^{ab}_{3} \dfrac{\partial W_{ab}}{\partial x^{a}_{3}}\right]}_{\text{divergence of the velocity difference}}}_{\text{pressure work}}\\ &+  \underbrace{2 \,\dfrac{\mu_{a}}{\rho_{a}} \left[\left(D'^{a}_{11}\right)^2+\left(D'^{a}_{22}\right)^2 +\left(D'^{a}_{33}\right)^2+\underbrace{2\left(D'^{a}_{12}\right)^2+2\left(D'^{a}_{23}\right)^2+2\left(D'^{a}_{13}\right)^2 }_{\text{because of symmetry of $ \bDprime $}}\right]}_{\text{energy dissipation due to viscous forces (viscous entropy)}}.
\end{align*}
In the function \verb@internal_force@ of SPHM, the term 
\[
  (p_{a}+p_{b}) \left[ v^{ab}_{1} \dfrac{\partial W_{ab}}{\partial x^{a}_{1}}+v^{ab}_{2} \dfrac{\partial W_{ab}}{\partial x^{a}_{2}}+v^{ab}_{3} \dfrac{\partial W_{ab}}{\partial x^{a}_{3}}\right]
\]
is stored in the variable \verb+p_work_part+.

This completes the presentation of the SPH approximation of Navier--Stokes equations.

\section{Other numerical aspects}\label{sec:art_visc_heat}
In this section, we present some other numerical aspects of the SPH method.

\subsection{Artificial viscosity}
The Monaghan type artificial viscosity was introduced in~\protect{\cite{Monaghan:1992}}, and reads
\[
   \Pi_{ab} = 
   \begin{cases}
      \dfrac{-\alpha \, c_{ab} \, \phi_{ab} + \beta \, \phi_{ab}^{2}}{\rho_{ab}}    &   \text{if}  \quad \vba \cdot \xba < 0, \\
      0   &   \text{if} \quad \vba \cdot \xba \geq 0,
   \end{cases}
\]
where $ \vba $ and $ \xba $ denote the relative velocity and the relative position, respectively, and
\[
   \phi_{ab} = \dfrac{h_{ab} \, \vba \cdot \xba}{\| \xba \|^{2} + (0.1 \, h_{ab})^{2} },
\]
where $ c_{ab} $, $ \rho_{ab} $, and $ h_{ab} $ are the average smoothing length, speed of sound, and density, respectively.
The parameters $ \alpha $ and $ \beta $ are usually taken equal to 1~\protect{\cite{Monaghan:1992}}. The term $ (0.1 \, h_{ab})^{2} $ prevents numerical singularities when two particles are approaching each other.

As it appears from its formulation, the term $ \Pi_{ab} $ is nonzero only in the case of material compression ($ \vba \cdot \xba < 0 $). It is added to the pressure terms in the SPH approximation of the momentum and energy equations.

\subsection{Artificial heat}
The artificial heat was introduced in~\protect{\cite{Noh:1987,Fulk:1994} with the following formulation
\[
   H_{a} = 2 \sum_{b=1}^{N} \dfrac{q_{ab}}{\rho_{ab}} \, \dfrac{e_{a} - e_{b}}{\| \xba \|^{2} + (0.1 \, h_{ab})^{2} } \, \xba \cdot \nabla_{a} W_{ab},
\]
where
\[
   q_{a} = \alpha \, h_{a} \, \rho_{a} \, c_{a} \, \left\lvert \nabla \cdot \bsym{v_{a}} \right\rvert + \beta \, h_{a}^{2} \, \rho_{a} \left(\nabla \cdot \bsym{v_{a}} \right)^{2},
\]
\[
   q_{b} = \alpha \, h_{b} \, \rho_{b} \, c_{b} \, \left\lvert \nabla \cdot \bsym{v_{b}} \right\rvert + \beta \, h_{b}^{2} \, \rho_{b} \left(\nabla \cdot \bsym{v_{b}} \right)^{2},
\]
and $ q_{ab} = \frac{1}{2} (q_{a} + q_{b}) $.
The term $ H_{a} $ is added to the SPH approximation of the energy equation.

\subsection{Variable smoothing length}
The early implementations of SPH \protect{\cite{Gingold:1977}} adopted the same smoothing length for all the particles and kept it fixed throughout the entire simulation. Later, problems involving fluid expansion and contraction required the introduction of a smoothing length varying in space according to the local density of the particles \protect{\cite{Liu:2003}}. Problems dealing with shocks pointed out the need of a smoothing length varying in both space and time.

Gingold and Monaghan \protect{\cite{Gingold:1982}} devised the simplest way to vary $h$ with time so that the number of neighboring particles of a given particle $a$ remains more or less the same. For each particle $a$ at a certain time instant, the smoothing length $h_{a}$ is given by\footnote{We drop the explicit dependence on $t$ for clarity.}
\begin{equation*}
   h_{a} = h_{0} \left( \dfrac{\rho_{0}}{\rho_{a}}\right)^{1/d},
\end{equation*}
where $h_{0}$ is the initial smoothing length, $\rho_{0}$ is the initial density, $\rho_{a}$ is the summation density at time $t$ for particle $a$, see eq.~\eqref{eq:summation_density}.
This proved to be a powerful way of calculating the smoothing length since it automatically makes $h$ varying in both space and time. The above expression has been improved by~\protect{\cite[eq.~(4.2)]{Monaghan:2005}}
\begin{equation*}
   h_{a} = 1.3 \left( \dfrac{m_{a}}{\rho_{a}}\right)^{1/d},
\end{equation*}
and further modified by the same author to avoid excessively large or small values of $h$
\begin{equation*}
   h_{a} = 1.3 \left( \dfrac{m_{a}}{A+\rho_{a}}\right)^{1/d},
\end{equation*}
where $A$ is a suitable constant. All these variants are implemented in~\verb+hsml_update+.

Another approach proposed by~\protect{\cite{Benz:1990}} consists in calculating $h$ from the rate of change of density, i.e.,
\begin{equation}\label{eq:h_evolving}
   \dfrac{\rmd h}{\dt}=-\dfrac{1}{d}\,\dfrac{h}{\rho}\,\dfrac{\rmd \rho}{\dt}.
\end{equation}
Let us recall the continuity equation~\eqref{eq:continuity}
\begin{equation*}
   \dfrac{\rmd \rho}{\dt}=-\rho \nabla\cdot \vel.
\end{equation*}
Substituting this expression into~\eqref{eq:h_evolving}, we obtain
\begin{equation*}
   \dfrac{\rmd h}{\dt} = \dfrac{h}{d} \, \nabla \cdot \vel,
\end{equation*}
which can be discretized using SPH approximation for the velocity divergence.

\subsection{Time step}
To perform the time integration, SPH makes use of an explicit numerical integration scheme. Explicit methods calculate the state of a system at a later time from the state of the system at the current time. For instance, given the equation
\begin{equation*}
   Y(t+\Delta t) = F(Y(t)),
\end{equation*}
where $ \Delta t $ is the time step, $F$ is the forcing term on the right-hand side, and $Y$ the function to be integrated, an explicit method finds $Y(t+\Delta t)$.

Explicit time integration schemes are subject to the Courant--Friedrichs--Lewy (CFL) condition for stability~\protect{\cite{Courant:1928,Liu:2003}}. The CFL condition requires the time step to be proportional to the smallest spatial particle resolution, which in SPH corresponds to the smallest smoothing length among all particles. In the SPH method, the CFL condition assumes the form~\protect{\cite{Monaghan:1989,Monaghan:1992}}
\begin{equation*}
   \Delta t = \min_{a}\left\lbrace \dfrac{h_{a}}{c_{a}} \right\rbrace,
\end{equation*}
where $c_{a}$ is the speed of sound for particle $a$.

\subsubsection{Adaptive time step}
As it can be seen from the above relationship, when the smoothing length gets smaller, the time step can become so small to be prohibitive, hence the need of a variable or \emph{adaptive time step}~\protect{\cite{Monaghan:1999,Gesteira:2010}}.
Different formulations for the adaptive time step have been proposed over the years. The most used ones are listed in the following.

The formulation of \protect{\cite{Monaghan:1989,Monaghan:1992}} involves the CFL condition, the artificial viscosity, and the force terms:
\begin{equation*}
   \Delta t = \min \left\lbrace 0.25 \Delta t_f, 0.4 \Delta t_{cv} \right\rbrace,
\end{equation*}
or, alternatively,
\begin{equation*}
   \Delta t = 0.3 \min \left\lbrace \Delta t_f, \Delta t_{cv}  \right\rbrace,
\end{equation*}
where $ \Delta t_f $ is the part of the time step due to external forces, and $\Delta t_{cv}$ is due to Courant and viscous forces.
The part of the time step due to external forces is \protect{\cite[p.~25]{Gesteira:2010}}
\begin{equation*}
   \Delta t_{f} = \min_{a} \left\lbrace \sqrt{\dfrac{h_{a}}{\| \fa \|}} \right\rbrace,
\end{equation*}
where $ \| \fa \| $ is the magnitude of the acceleration vector for particle $a$.
The part of the time step due to Courant and viscous forces is
\begin{equation*}
   \Delta t_{cv} = \min_{a} \left\lbrace \dfrac{h}{c_{s}+\displaystyle\max_{b}\left|\dfrac{h\vba\cdot\,\xba}{\| \xba \|^{2}}\right|} \right\rbrace.
\end{equation*}
An alternative formulation of $\Delta t_{cv}$ is
\begin{equation*}
   \Delta t_{cv} = \min_{a} \left\lbrace \dfrac{h_{a}}{c_{a} + 0.6 \, (\alpha_{\pi}c_{a}+\beta_{\pi}\displaystyle\max_{k} \left\lbrace \phi_{ab} \right\rbrace ) } \right\rbrace,
\end{equation*}
where $ \alpha_{\pi} $, $ \beta_{\pi} $, and $ \phi_{ab} $ are the terms related to artificial viscosity (see section~\ref{sec:art_visc_heat}), and $ k $ is the number of interacting pairs.

\section{Boundary treatment}\label{sec:boundary_treatment}
\epigraph{\emph{SPH has been unable to treat generalized boundary conditions, but this is not an inherent limitation of the method.}}{--- Randles and Libersky, 1996}
Since the birth of SPH, several boundary treatments have been proposed. In the early computational fluid dynamics uses of SPH, simple boundary conditions such as non-penetrating surfaces were adopted. Nonetheless, being a particle method, the boundary of the simulation domain $ \Omega $ is never well defined.
The problem is that close to the boundaries of the simulation domain, the SPH method is affected by particle deficiency, since the integral of the kernel function is truncated by the boundary. This situation is illustrated in Figure~\ref{fig:SPH_near_boundary}.

\begin{figure}[H]
\centering
\includegraphics[width=0.45\textwidth]{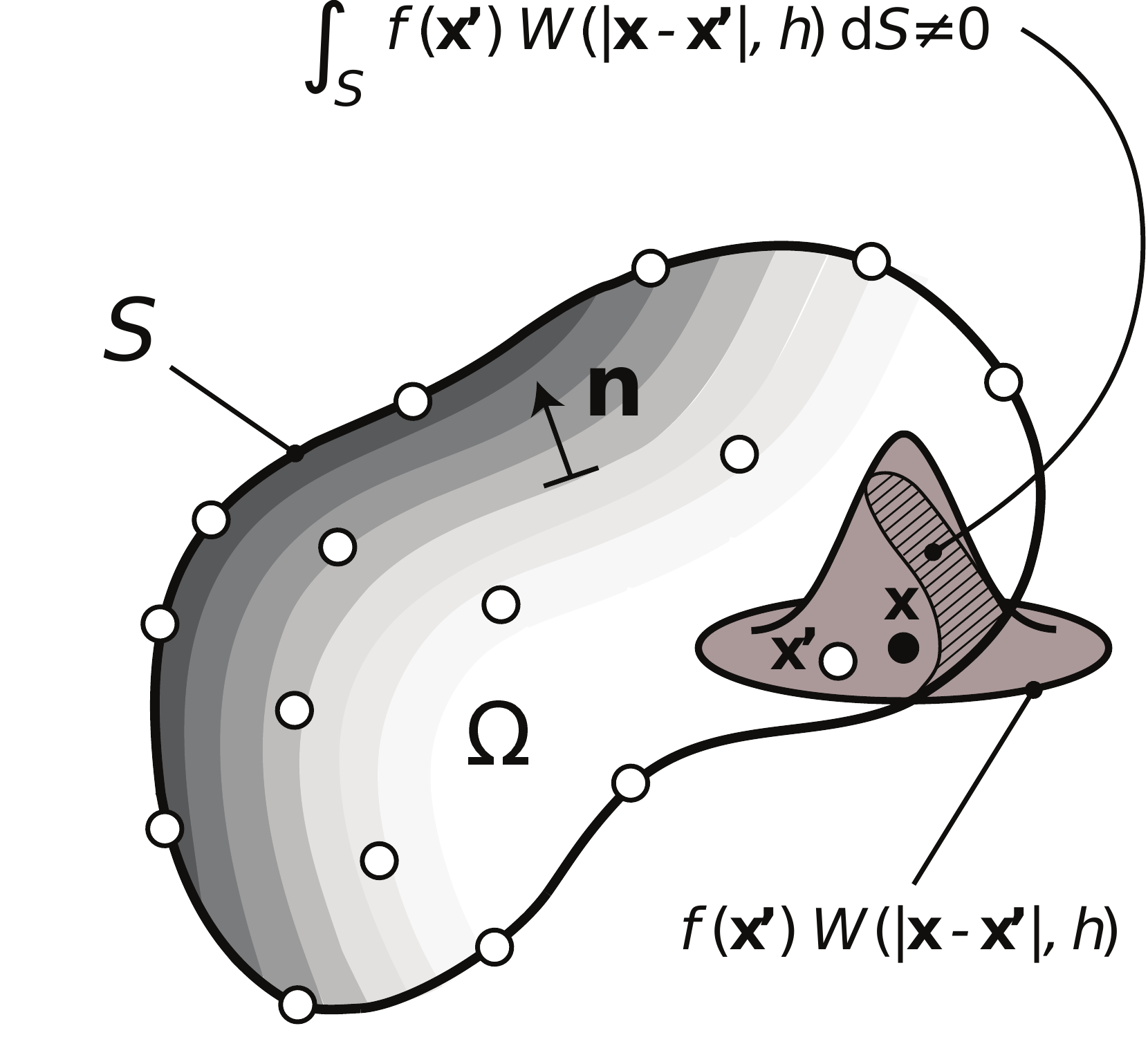} 
\caption{Behavior of SPH near the boundaries.}
\label{fig:SPH_near_boundary}
\end{figure}

In order to be stable, the SPH method needs a sufficient and necessary number of particles within the support domain of $ \kappa h $, where $ \kappa $ is a coefficient to be specified. In one, two, and three dimensions the number of neighboring particles (including the particle itself) should be about 5, 21, and 57, respectively, if the particles are arranged in an initial lattice with a smoothing length of 1.2 times the particle spacing $ \Delta x $, and $\kappa = 2$ \protect{\cite{Liu:2003}}. Figure~\ref{fig:stability} illustrates this concept for the two-dimensional case.

\begin{figure}[H]
\centering
\includegraphics[width=0.35\textwidth]{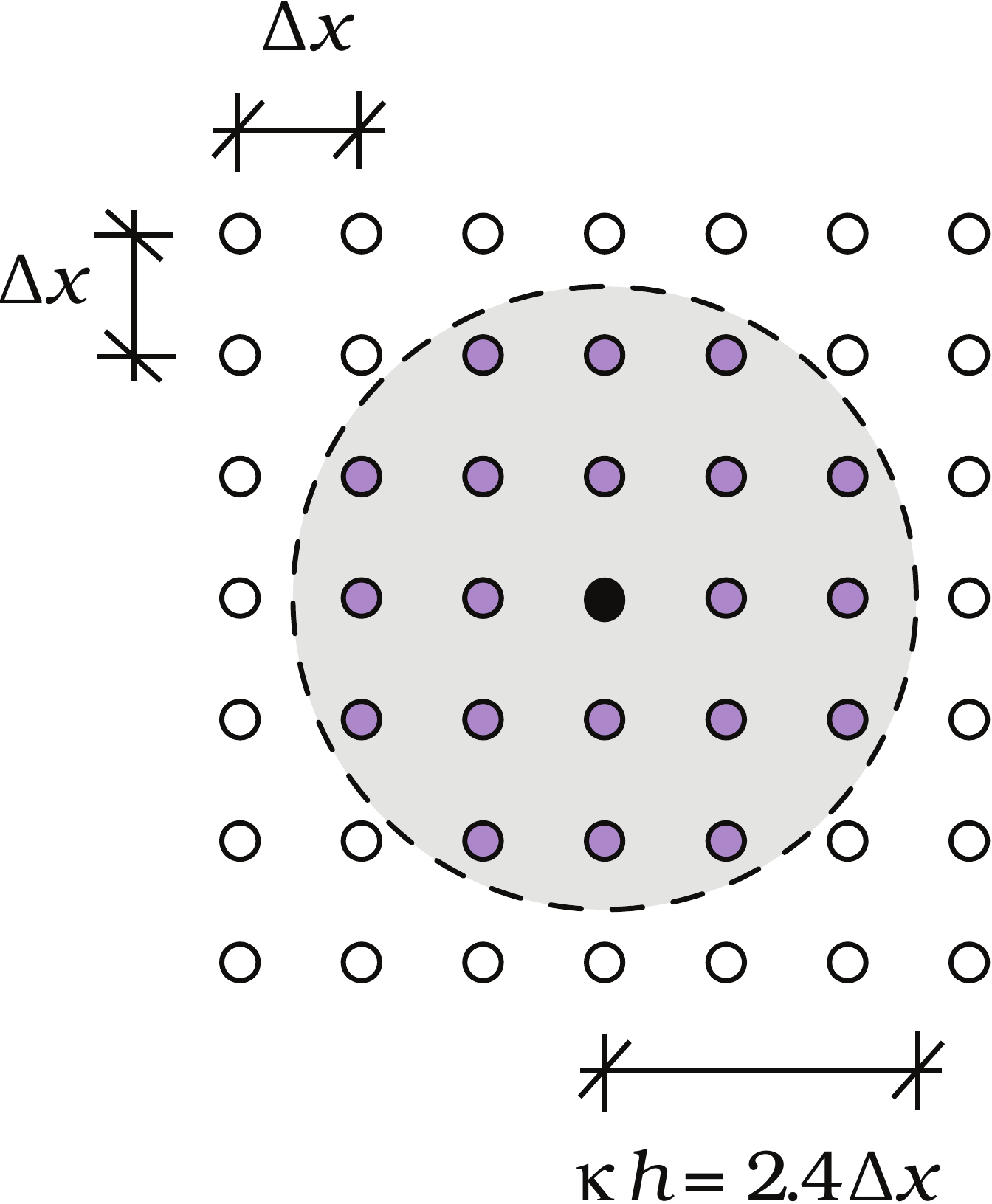} 
\caption{Stability of two-dimensional SPH.}
\label{fig:stability}
\end{figure}

Therefore, for particles near or on the boundary, only particles inside the domain contribute to the summation, and this one-sided contribution leads to wrong solutions.

Over the years, several authors have proposed different solutions to address the problem of boundary treatment in SPH. In all of these approaches, boundaries are typically defined by lines of particles that exert repulsive forces on fluid particles. In this section, we explore the following concepts and formulations:
\begin{itemize}
\item Ghost particles~\protect{\cite{Libersky:1993}};
\item Lennard--Jones potential for repulsive boundary forces~\protect{\cite{Monaghan:1994}};
\item Normal boundary forces~\protect{\cite{Monaghan:1999,Monaghan:2003}};
\item Radial boundary forces~\protect{\cite{Monaghan:2009}}.
\end{itemize}

\subsection{Ghost particles}
One of the first approaches for boundary treatment is the introduction of \emph{ghost particles}.
In~\protect{\cite{Libersky:1993}}, Libersky was the first to introduce ghost particles to reflect a symmetrical surface, and in~\protect{\cite{Monaghan:1994}} Monaghan used a line of virtual particles located right on the solid boundary, to produce a \emph{highly repulsive force} to the particles near the boundary, in order to avoid penetration.

\protect{\cite{Liu:2003}} merged and improved these two approaches to treat the solid boundary conditions. According to~\protect{\cite{Liu:2003}}, virtual particles (VP) can be of two types:
\begin{itemize}
\item Type I: right on the solid boundary (similar to \protect{\cite{Monaghan:1994}});
\item Type II: they fill in the region outside the boundary and close to it (similar to \protect{\cite{Libersky:1993}}).
\end{itemize}

Virtual particles of type I take part in kernel and particle approximation for the real particles, but their position and physical variables do not evolve in the simulation. They are used to exert a repulsive boundary force to prevent the interior particles from penetrating the solid boundary.

The ghost particles (i.e., VP of type II) can be applied to both \emph{solid boundary} and \emph{free surfaces}. They are constructed as follows. If a real particle $ a $ is located within the distance $\kappa h$ from the boundary, then a VP is symmetrically placed on the outside of the boundary (see Figure~\ref{fig:VP_construction}). These particles have the \emph{same density} and \emph{pressure} as the corresponding real particles, but \emph{opposite velocity}. VP of type II do not evolve their parameters, since they are created symmetrically to the corresponding real particles at every time step.

\begin{figure}[htbp]
   \centering
   \includegraphics[width=0.70\textwidth]{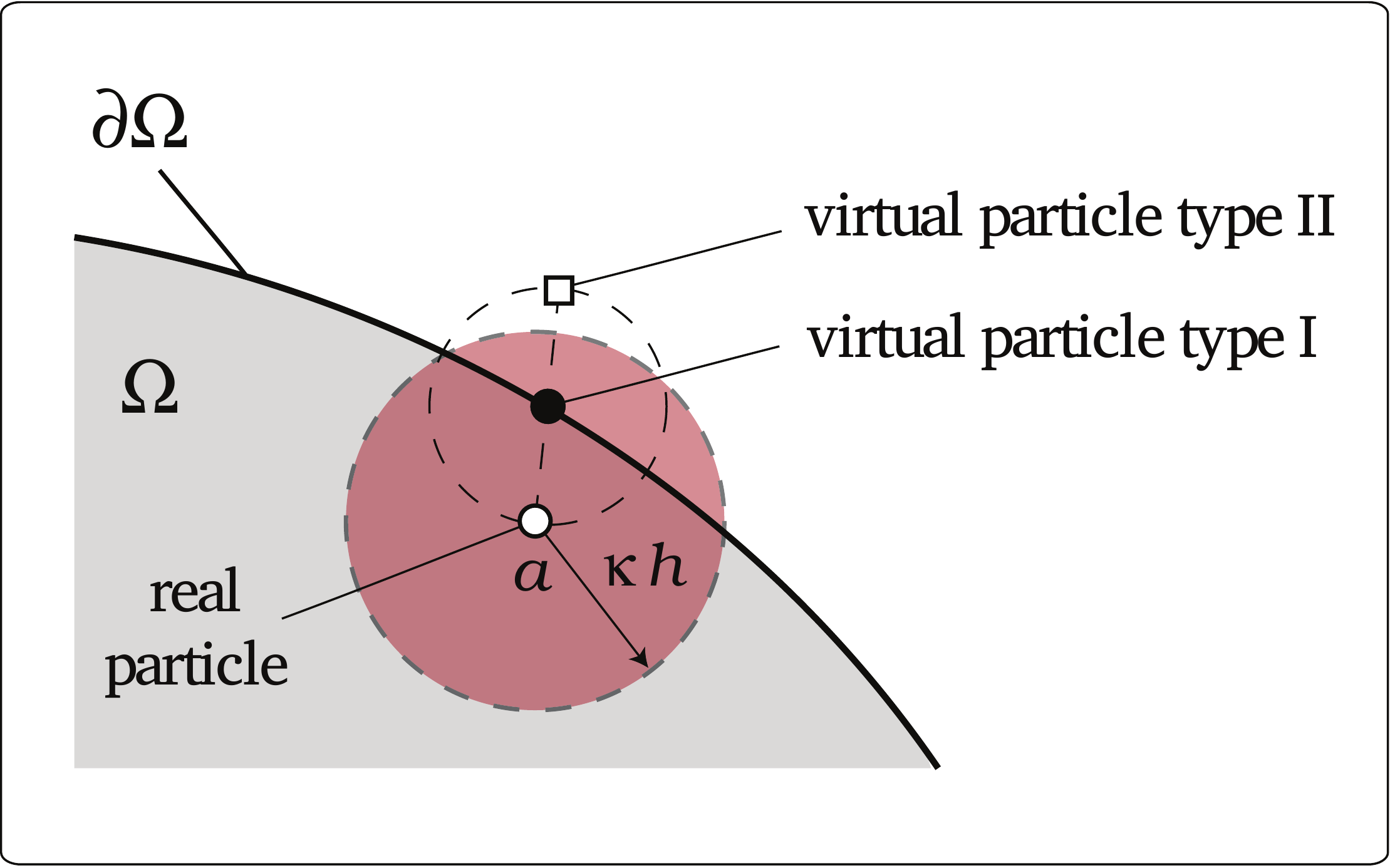} 
   \caption{Construction of VP of type II according to \protect{\cite{Liu:2003}}.}
   \label{fig:VP_construction}
\end{figure}
\begin{figure}[htbp]
   \centering
   \includegraphics[width=0.70\textwidth]{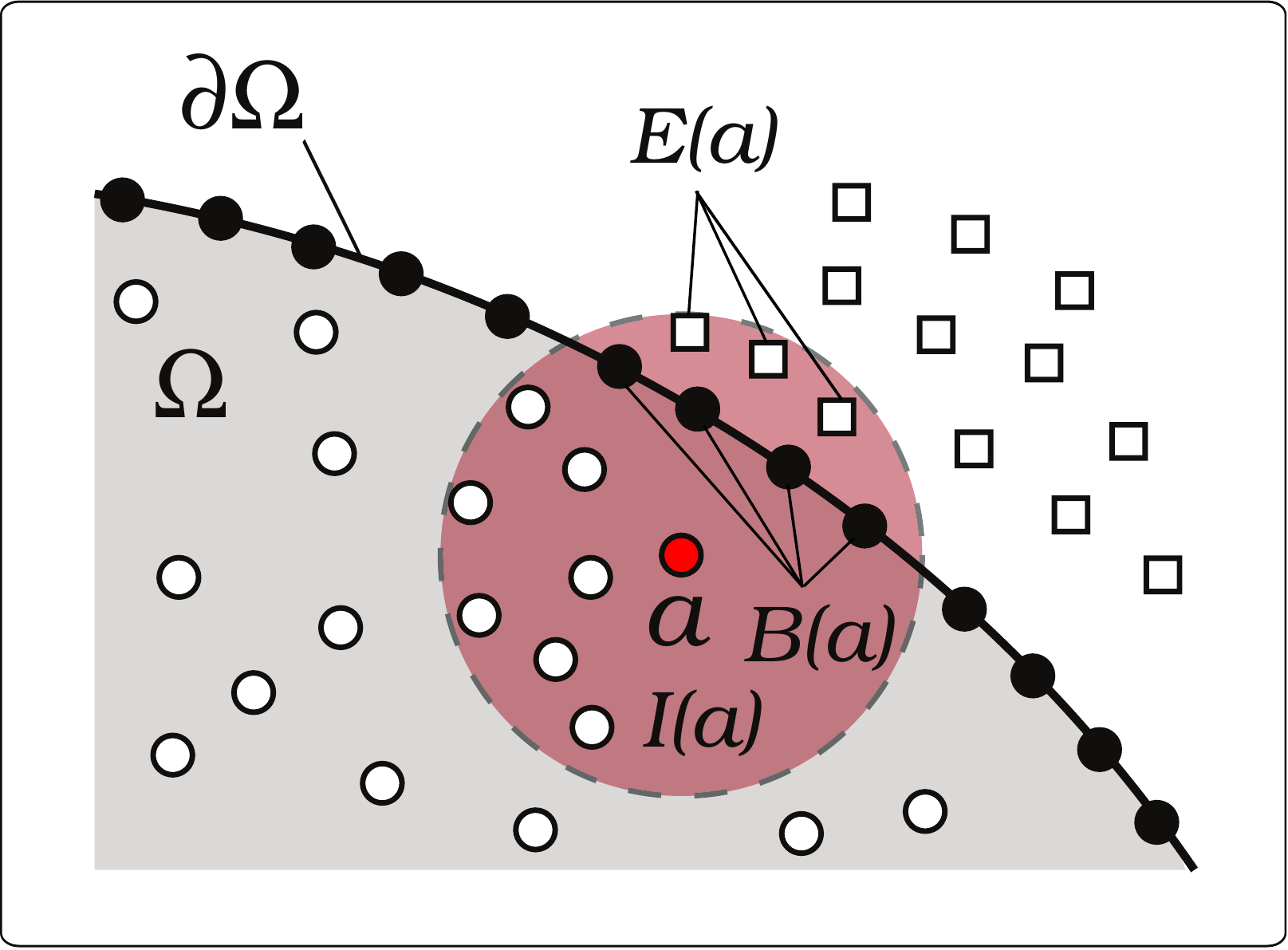} 
   \caption{Types of particles falling into the support domain of a particle $a$ close to the boundary.}
   \label{fig:particle_types}
\end{figure}

Summarizing, for a particle $ a $ close to the boundary, the set of particles falling into its domain of influence can be divided into three subsets (see Figure~\ref{fig:particle_types}):
\begin{itemize}
\item $I(a)$: interior or real particles;
\item $B(a)$: boundary particles, or virtual particles of type I;
\item $E(a)$: exterior particles, or virtual particles of type II (ghost particles).
\end{itemize}

The total number of particles $N$ in the support domain of a particle $a$ close to the boundary is therefore $ N(a) = I(a) \cup B(a) \cup E(a) $.

\subsubsection{Pros and contras}
This approach has the advantages to restore the SPH consistency near the boundaries and to prevent non-physical penetration through the solid boundary. Yet in the original presentation of \protect{\cite{Randles:1996}}, a very important point was missing: \emph{how to generate} the ghost particles. Typically, the ghost particles are generated by mirroring the fluid particles, and this implies that they must adapt with the fluid particles at each time step, causing an additional computational effort. The ghost particles become particularly unwieldy in presence of corners or surfaces with high curvature, since in these situations they cannot be placed without ambiguity.

\subsection{Lennard--Jones potential} \label{sec:Lennard-Jones}
The formulation of this repulsive force is based on the known forces between molecules \protect{\cite{Monaghan:1994}. In fact, it takes the form of a Lennard--Jones potential.

An example of Lennard--Jones potential is shown in Figure~\ref{fig:12-6-Lennard-Jones-Potential}.

\begin{figure}[htbp]
\centering
\includegraphics[width=0.63\textwidth]{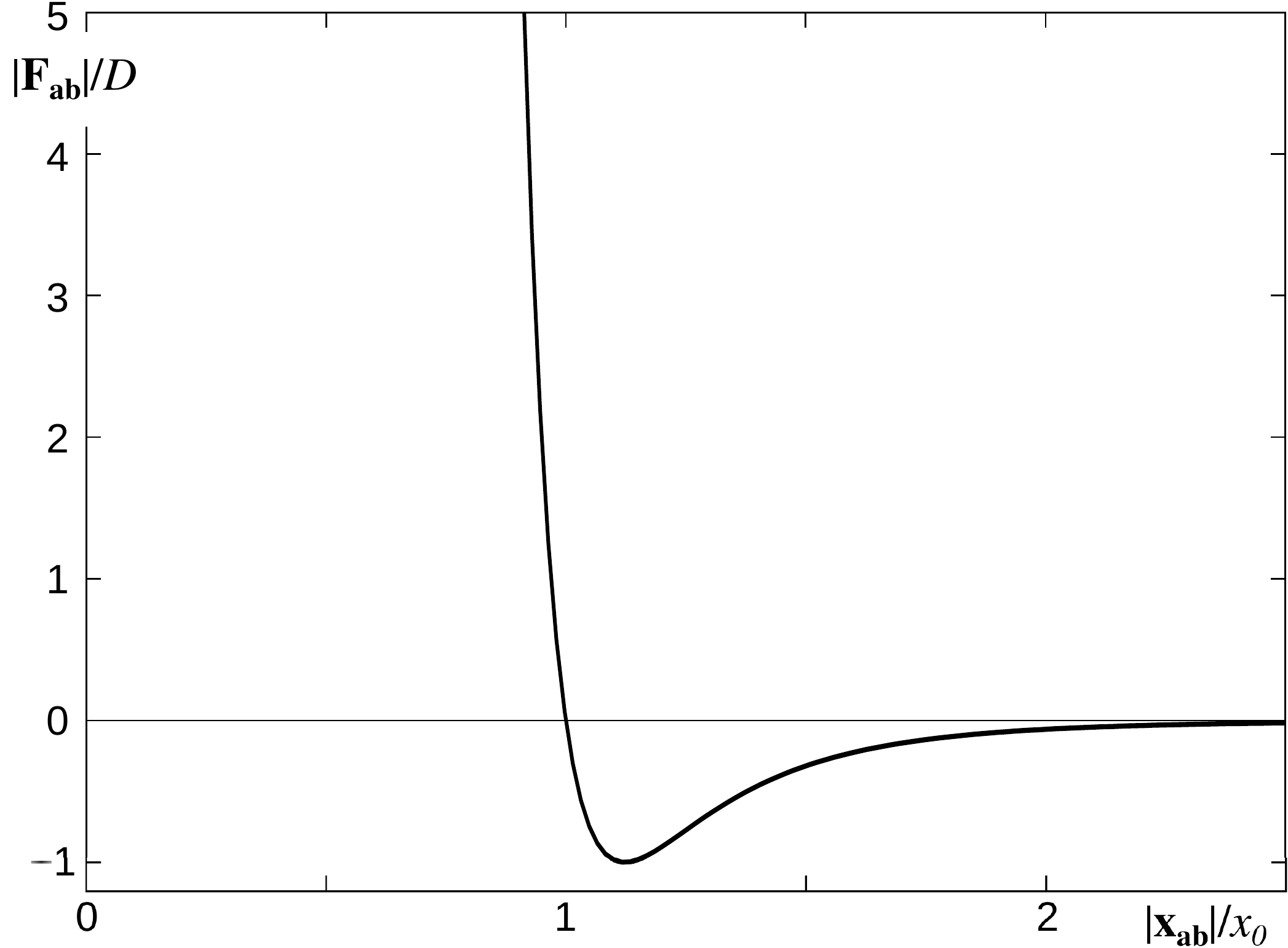} 
\caption{A graph of strength versus distance for the 12-6 Lennard--Jones potential.}
\label{fig:12-6-Lennard-Jones-Potential}
\end{figure}

The way this potential works is quite simple. If a real particle $ a $ is approaching a boundary particle $ k $, then a pairwise repulsive force is applied along the centerline of these two particles. The mathematical formulation is
\begin{equation*}
\fak =
\begin{cases}
D \left[ \left( \dfrac{x_{0}}{\| \xak \|}\right)^{n_1} -\left( \dfrac{x_{0}}{\| \xak \|}\right)^{n_2} \right]\,\dfrac{\xak}{\| \xak \|^2}, &\qquad \| \xak \|\leq x_{0}, \\
0, &\qquad \| \xak \| > x_{0},
\end{cases}
\end{equation*}
where $\| \xak \|$ is the distance between particle $a$ and boundary particle $k$, $x_{0}$ is a cutoff distance, $D$ is a problem parameter, and usually $n_1 = 12$ and $n_2 = 6$. 
The coefficient $D$ should be chosen considering the physical configuration. For problems involving dams, bores, weirs with fluid depth $ H $, we may set $ D = 5gH $, but also $ D = 10gH $ or $ D = gH $. 

The cutoff distance $x_{0}$ is usually selected to be approximately equal to the initial particle spacing $\Delta x$. If it is too large, then some particles may feel the repulsive force in the initial distribution already. If it is too small, then the particles will penetrate the boundary before feeling the repulsive force.

\subsubsection{Pros and contras}
If the parameters are well tuned, the Lennard--Jones potential formulation avoids particle penetration into the boundary, ensuring that the velocity component normal to the boundaries vanishes. Nonetheless, Lennard--Jones forces are not satisfactory, since a particle moving parallel to the boundary is subject to a non-uniform normal force and a non-zero tangential force, leading to large disturbances in the flow near a boundary. See, for example, the shear-driven cavity problem in section~\ref{sec:shear_cavity}. To avoid this shortcoming, some authors adopted boundary particle forces based on an interpolation procedure that we are going to present in the next section.

\subsection{Boundary force approach}\label{sec:boundary_force}
According to \protect{\cite{Monaghan:2003}}, ``\emph{the force per unit mass }$\fk$ \emph{on the boundary particles is due to the fluid particles unless the moving rigid body strikes a fixed boundary}''. Neglecting the latter case for simplicity, we can write
\begin{equation*}
   \fk=\sum_{a}\fka,
\end{equation*}
where $\fka$ represents the force per unit mass on boundary particle $k$ due to fluid particles $a$.
The force $\fk$ on each boundary particle is computed by summing up the contributions from all the surrounding water particles which fall within the supporting kernel. 

In this section, we will review two different formulations available in literature for the calculation of the boundary force $\fka$: the normal force and the radial force approaches.

\subsubsection{Normal force} \label{sec:normal_boundary_force}
\protect{\cite{Monaghan:2003}} proposed that the forces from neighboring boundary particles should give rise to a force normal to the boundary. Let's consider the $k$-th boundary particle and the fluid particle $a$, as shown in Figure~\ref{fig:normal_force_approach}.

\begin{figure}[htbp]
   \centering
   \includegraphics[width=0.45\textwidth]{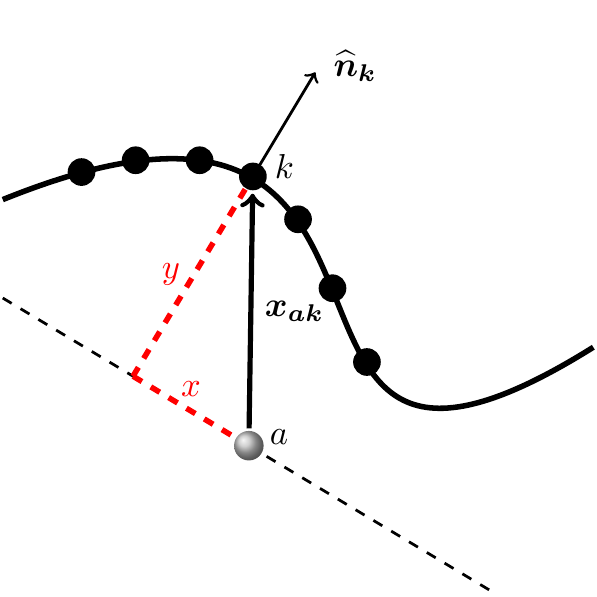} 
   \caption{Normal boundary force approach.}
   \label{fig:normal_force_approach}
\end{figure}

Then the repulsive force per unit mass on boundary particle $k$ due to fluid particle $a$ is given by
\begin{equation*}
   \fka=-\dfrac{m_{a}}{m_{a}+m_{k}}\,B(x,y)\,\nk,
\end{equation*}
and, in turn, because of Newton's third law, the force per unit mass on $a$ due to $k$ is
\begin{equation}\label{eq:normal_force_approach}
   \fak = \dfrac{m_{k}}{m_{a}+m_{k}}\,B(x,y)\,\nk.
\end{equation}
Let us have a closer look to the parameters involved in these equations:
\begin{itemize}
\item $\nk$ is the outward unit normal to the boundary at the location of particle $k$;
\item $m_{a}$ and $m_{k}$ are the masses of particles $a$ and $k$, respectively;
\item $x$ and $y$ are the tangential and the normal distance between $a$ and $k$, respectively;
\item $B(x,y)$ is a function of the local coordinates $x$ and $y$.
\end{itemize}
The normal distance $y$ can be viewed as the projection on the direction of $\nk$ of the vector distance $ \xak = \xk -\xa $ between $a$ and $k$, thus it can be obtained from the scalar product
\begin{equation*}
   y=\xak\cdot \nk.
\end{equation*}
The total force per unit mass on fluid particle $a$ due to all the boundary particles $k$ in the support domain $ B(a) $ is
\begin{equation}\label{eq:total_force_per_unit_mass}
   \fa = \sum_{k\in B(a)} \fak.
\end{equation}
This force is then added to the SPH approximation of the equations of motion~\eqref{eq:SPH_approx_eom} with the first form of symmetrization
\begin{equation}\label{eq:SPH_approx_eom_plus_boundary_force}
\dfrac{\dv_{i}^{a}}{\dt}=-\sum_{b=1}^{N}m_{b}\left( \dfrac{p_{a}}{\rho^{2}_{a}}+\dfrac{p_{b}}{\rho^{2}_{b}}\right) \dfrac{\partial W_{ab}}{\partial x_{i}^{a}} + 2 \sum_{b=1}^{N} m_{b} \left( \dfrac{\mu_{a} D'^{a}_{ij}}{\rho^{2}_{a}}+\dfrac{\mu_{b} D'^b_{ij}}{\rho^{2}_{b}}\right) \dfrac{\partial W_{ab}}{\partial x_{j}^{a}} + f_{i}^{a},
\end{equation}
with $ i = 1, \ldots, d $.

\subsubsection{Choice of the function \texorpdfstring{$ B(x,y) $}{}}
We still have to define the form of the function $B(x,y)$ appearing in \eqref{eq:normal_force_approach}. In general, the variation of $B(x,y)$ with $x$ should ensure that the force on a fluid particle moving parallel to the boundary is constant. Moreover, $B(x,y)$ should be chosen so that it rapidly increases as $y$ tends to zero, i.e., as the fluid particle is approaching the boundary, to prevent penetration.

\paragraph{Choice 1.} \protect{\cite{Monaghan:2003}} wrote $B(x,y)$ as a separable variable function
\begin{equation*}
B(x,y)=\Gamma(y)\,\chi(x),
\end{equation*}
where
\begin{equation*}
\chi(x)=\begin{cases}
\left(1-\dfrac{x}{\Delta p} \right) & \text{if} \quad 0<x<\Delta p,\\
0 & \text{otherwise},
\end{cases}
\end{equation*}
with $\Delta p$ being the spacing between the boundary particles\footnote{Typically the spacing between the boundary particles $\Delta p$ is about one half of the initial fluid particle spacing $\Delta x$ \protect{\cite{Monaghan:2009}}.}.The function $\chi(x)$ ensures that a fluid particle moving parallel to the boundary will experience a constant force.
The function $\Gamma(y)$ has a form related to the gradient of the kernel \protect{\cite{Monaghan:2003,Monaghan:2005}}, i.e.,
\begin{equation*}
\Gamma(y)=\begin{cases}
\frac{2}{3}\,\beta & \text{if} \quad 0<q<\frac{2}{3},\\
\beta \left(2q - \frac{3}{2}\,q^2 \right)  & \text{if} \quad \frac{2}{3}<q<1,\\
\frac{1}{2}\,\beta \left( 2-q \right)^2 & \text{if} \quad 1<q<2,\\
0 & \text{otherwise},
\end{cases}
\end{equation*}
where
\begin{itemize}
\item  $q=y/h$, with $ y $ being the normal distance (see Figure~\ref{fig:normal_force_approach});
\item $\beta=0.02\,c_{s}^{2}/y$, being $c_{s}$ the speed of sound. This term represents an estimate of the maximum force per unit mass necessary to stop a particle moving at the estimated maximum speed. The factor $1/y$ ensures that a faster moving particle can be stopped.
\end{itemize}
Overall, the function $\Gamma(y)$ quickly increases as $y$ decreases in order to prevent the fluid particle from penetrating the boundary of the simulation domain.

\paragraph{Choice 2.} A similar approach was given by~\protect{\cite{Monaghan:1999}}, who defined $\fak$ as
\begin{equation*}
   \fak=R(y)\,P(x)\,\nk.
\end{equation*}
The specific form of $R(y)$ is not crucial, however, as before, it should rapidly increase as $y$ tends to zero. The authors proposed the following formulation:
\begin{equation*}
R(y)=\begin{cases}
A\,\dfrac{1}{\sqrt{q}}\,(1-q) \quad &\text{if} \quad q<1,\\
0 & \text{otherwise},
\end{cases}
\end{equation*}
where
\begin{itemize}
\item $q=y/2h$;
\item $A$ is a coefficient having dimensions of an acceleration:
\begin{equation*}
A = \dfrac{1}{h}\, 0.01 \, c_{a}^2.
\end{equation*}
\end{itemize}

As before, the function $P(x)$ is designed so that a fluid particle moving parallel to the boundary will experience a constant repulsive boundary force
\begin{equation*}
P(x)=
\begin{cases}
\dfrac{1}{2} \left( 1+ \cos \dfrac{\pi x}{\Delta p}\right) \quad & \text{if}\quad x<\Delta p,\\
0  & \text{otherwise}.
\end{cases}
\end{equation*}

\paragraph{Choice 3.}~\protect{\cite{Gesteira:2010}} introduced a correction factor $ \epsilon(z,u_{\perp}) $ into the original formulation of~\protect{\cite{Monaghan:1999}}
\begin{equation*}
   \fak=R(y)\,P(x)\, \epsilon(z,u_{\perp}) \,\nk.
\end{equation*} 
The function $ \epsilon(z,u_{\perp}) $ is used to adjust the magnitude of the repulsive force according to the local water depth and velocity of the fluid particle normal to the boundary.
Please note that this is the formulation that has been implemented in the SPHysics code, according to the guide of~\protect{\cite{Gesteira:2010}}.

\subsubsection{Pros and contras}
The main advantage of using a normal boundary force approach over the Lennard--Jones repulsive forces is that fluid particles moving parallel to the boundary experience a constant repulsive boundary force. This, as we have seen, is one flaw of the Lennard--Jones repulsive forces which is here avoided.
Unfortunately, the problem with all the normal boundary force approaches lies in the calculation of the outward normals to a surface. In fact, this is not an easy task to be achieved in SPH. Sometimes the calculation of the outward unit normal may be ambiguous, as illustrated in Figure~\ref{fig:normal_to_boundary}.

\begin{figure}[htbp]
\centering
\includegraphics[scale=0.6]{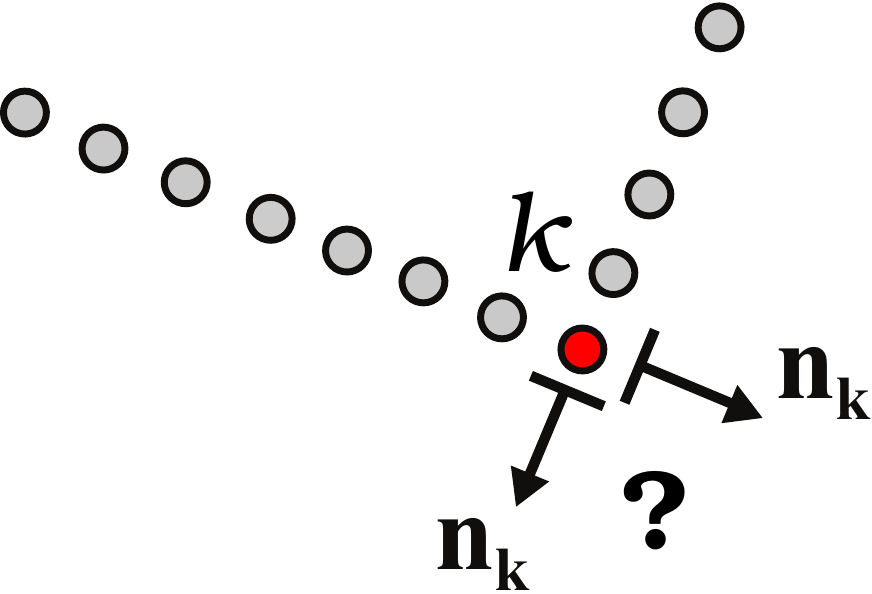} 
\caption{Ambiguity in the calculation of the outward normal at a sharp corner.}
\label{fig:normal_to_boundary}
\end{figure}

In this example, a corner particle has two outward normals, so it may interact with the fluid particles using either one or another of the normals to the lines that make up the corner, or we may assign a normal with direction halfway between the normals to the two straight lines. In any case, it is clear that, because of this ambiguity in defining the outward normal, we have to treat the corner particles differently from the other boundary particles, hence we need to allow the code to identify corner particles. This is not desirable since we it inevitably increases the computational effort of the simulation.

\subsubsection{Radial force}
In order to solve the problem of treating the corner particles differently from the other boundary particles,~\protect{\cite{Monaghan:2009}} suggested to adopt repulsive boundary forces with radial direction. See Figure~\ref{fig:radial_force_approach} for an illustration.

\begin{figure}[htbp]
   \centering
   \includegraphics[width=0.65\textwidth]{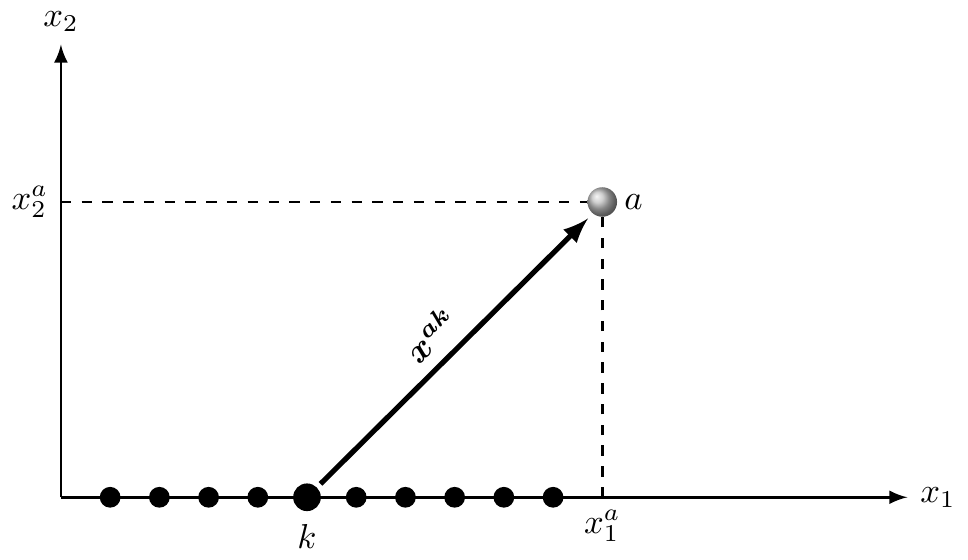} 
   \caption{Radial boundary force approach.}
   \label{fig:radial_force_approach}
\end{figure}

The proposed formulation for $\fak$ is
\begin{equation}\label{eq:radial_boun_force}
   \fak=\dfrac{K}{\beta}\,\dfrac{\xak}{\| \xak \|^2}\,W(\| \xak \|/h)\,\dfrac{2m_{k}}{m_{a}+m_{k}},
\end{equation}
where
\begin{itemize}
\item $K=gD$, with $ D $ being the initial water depth;
\item $\beta = \Delta x/\Delta p $. This parameter ensures that if we change the spacing between the boundary particles, the force on the fluid does not change.
\end{itemize}
If we assume that all particle masses are equal, the above equation~\eqref{eq:radial_boun_force} reduces to
\begin{equation}\label{eq:radial_boun_force2}
   \fak=\dfrac{K}{\beta}\,\dfrac{\xak}{\| \xak \|^2}\,W(\| \xak \|/h).
\end{equation}
We then proceed as in section~\ref{sec:normal_boundary_force} using \eqref{eq:total_force_per_unit_mass} and \eqref{eq:SPH_approx_eom_plus_boundary_force}.

This approach removes the need of computing the outward unit normals to the boundaries, which as we have seen was the origin of some ambiguities in the normal boundary force formulation. Moreover, it simplifies SPH algorithms and turns out to be superior to other formulations when dealing with complicated boundaries.

\section{Simulations}\label{sec:simulations}
In this section, we present the numerical results for two classical benchmark problems: the one-dimensional shock tube~\protect{\cite[p.~156]{Liu:2003}} and the two-dimensional shear-driven cavity problem~\protect{\cite[p.~94, Example 3.6]{Liu:2003}}.

We conducted our experiments on a laptop Lenovo ThinkPad T460s with Ubuntu 22.04.1 LTS and MATLAB R2022a installed, with Intel Core i7-6600 CPU, 20GB RAM, and Mesa Intel HD Graphics 520.

%
%
%
%
%
%
%
%

\subsection{Sod shock tube problem}\label{sec:sod_shock_tube}
The Sod shock tube~\protect{\cite[\S3]{Sod:1978}} is a long straight tube filled with gas, separated by a membrane into two parts of different pressures and densities.
The sudden removal of the membrane generates a shock wave, a rarefaction wave, and a contact discontinuity. The shock wave moves into the region with lower density, the rarefaction wave moves into the region with higher density, and the contact discontinuity forms near the center and moves into the low-density region following the shock wave.

This is a good benchmark because it has an exact solution available, and it has also been implemented in SPHM. The script \verb@Driver_Sod_shocktube@ can be used to generate the Sod shock tube profiles using Sod's original data.

In the simulation of this section, the initial conditions have been chosen as in~\protect{\cite{Monaghan:1983,Hernquist:1989}}, and are as follows: 
\[
   x \in [-0.6, \, 0.0], \quad \rho = 1, \quad \vel = 0, \quad e = 2.5, \quad p = 1, \quad \Delta x = 0.001875.
\]
\[
   x \in [0.0, \, 0.6], \quad \rho = 0.25, \quad \vel = 0, \quad e = 1.795, \quad p = 0.1795, \quad \Delta x = 0.0075.
\]
We use 400 particles in the simulation, all having the same mass $ m = 0.001875 $. The shock tube length is 1.2 meters, corresponding to the interval $ [-0.6, \, 0.6] $ on the real axis. In the left subinterval $ [-0.6, \, 0.0] $ we place 320 evenly distributed particles, while in the right subinterval $ [0.0, \, 0.6] $ we place 80 evenly distributed particles. The reason for this distribution is to have a discontinuous density profile along the tube.
We adopt the quartic smoothing function from~\protect{\cite[p.~92]{Liu:2003}}.
Moreover, we use the equation of state~\eqref{eq:eos_gas} for ideal gas, with $ \gamma = 1.4 $. We employ a fixed time step $ \Delta t = 0.005 $ s, and the simulation runs for a total 40 time steps, so that the final time of the simulation is $ t = 0.20 $ s.

The Monaghan type artificial viscosity introduced in~\protect{\cite{Monaghan:1992}} and presented in section~\ref{sec:art_visc_heat} is used to resolve the shock front.

To perform this simulation and reproduce the results presented in this section, simply run the script \verb+Driver_SPHM+ with the parameter \verb+sph.example = 1+. This will generate the mfiles in the results folder. Then, run the script \verb+Driver_shocktube_profiles+ to plot the profiles in Figure~\ref{fig:shocktube_profiles_1}. 

\begin{figure}[htbp]
    \centering
    \begin{minipage}{0.48\textwidth}
        \centering
        \includegraphics[width=\textwidth]{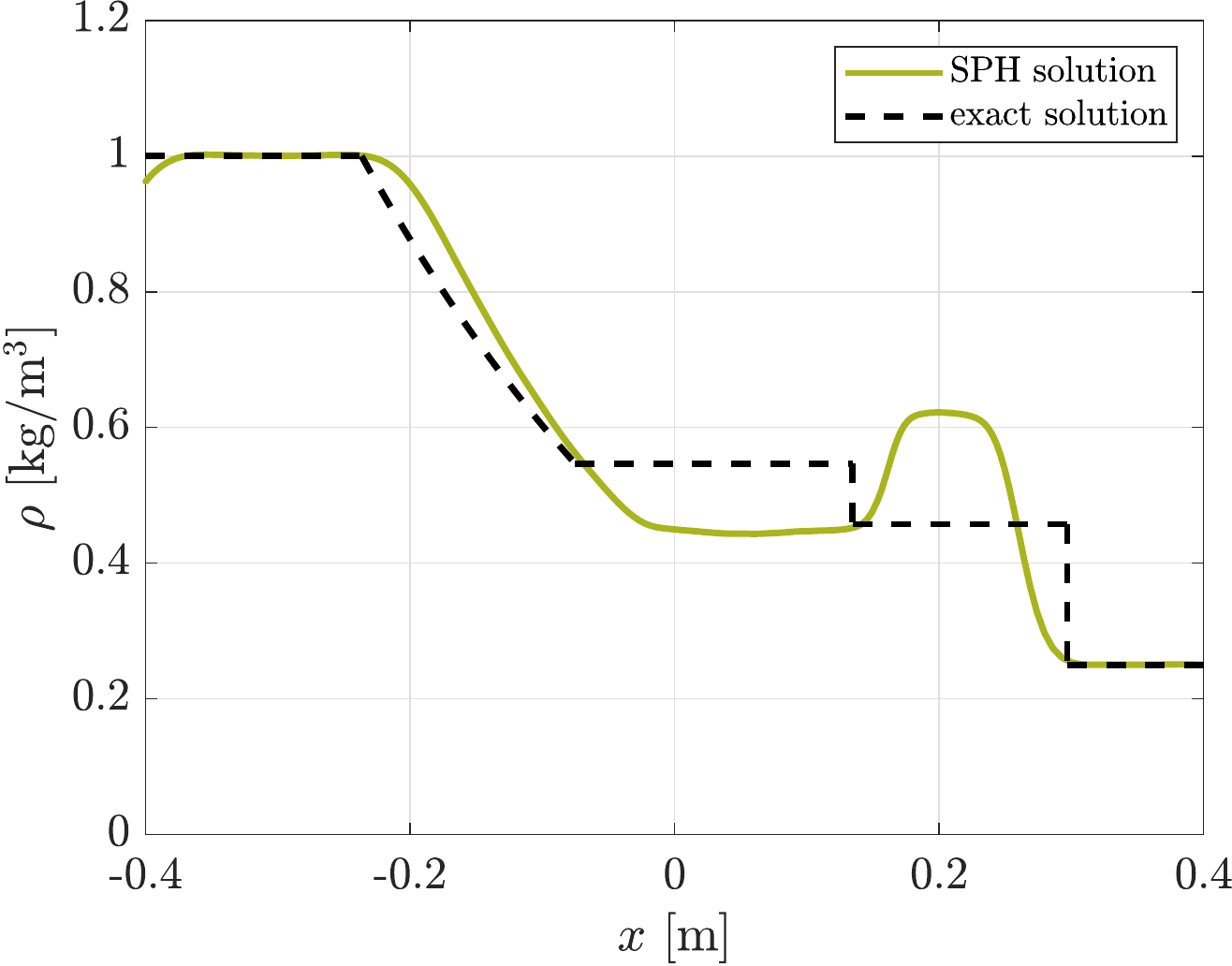}
        {\scriptsize (a) \emph{Density profile}}
    \end{minipage}\hfill
    \begin{minipage}{0.48\textwidth}
        \centering
        \includegraphics[width=\textwidth]{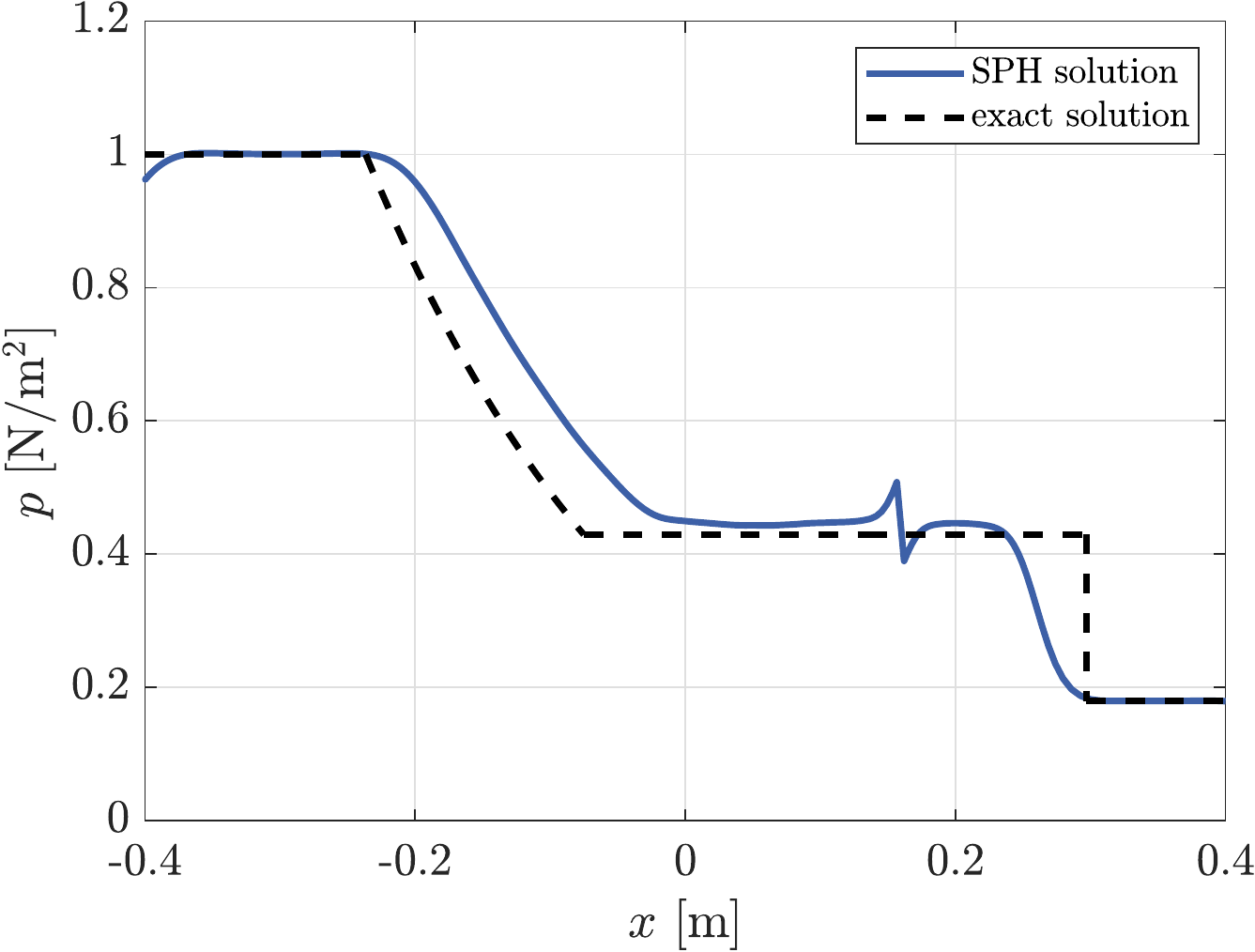}
        {\scriptsize (b) \emph{Pressure profile}}
    \end{minipage}
\end{figure}

\begin{figure}[htbp]
    \centering
    \begin{minipage}{0.48\textwidth}
        \centering
        \includegraphics[width=\textwidth]{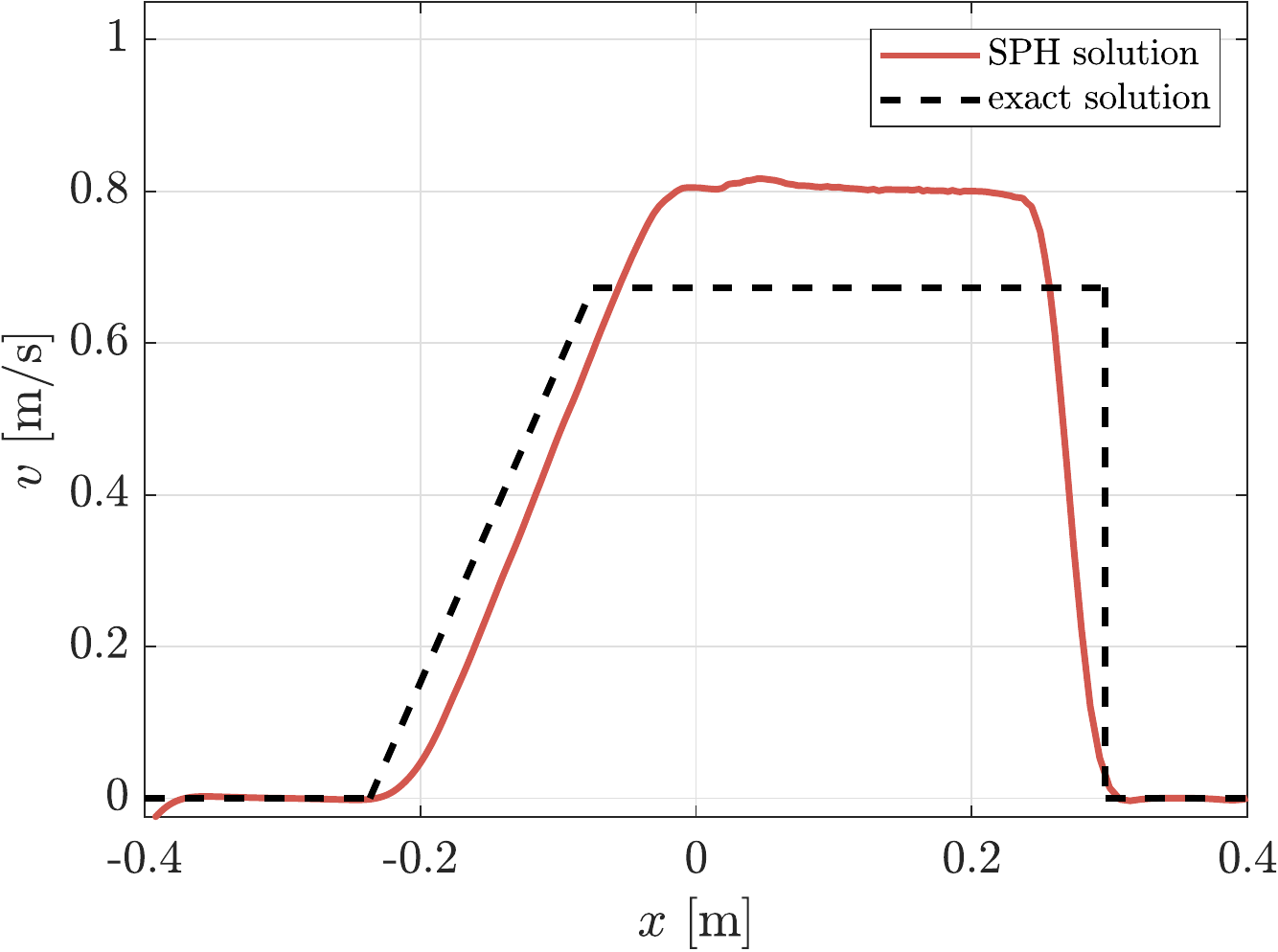}
        {\scriptsize (c) \emph{Velocity profile}}
    \end{minipage}\hfill
    \begin{minipage}{0.48\textwidth}
        \centering
        \includegraphics[width=\textwidth]{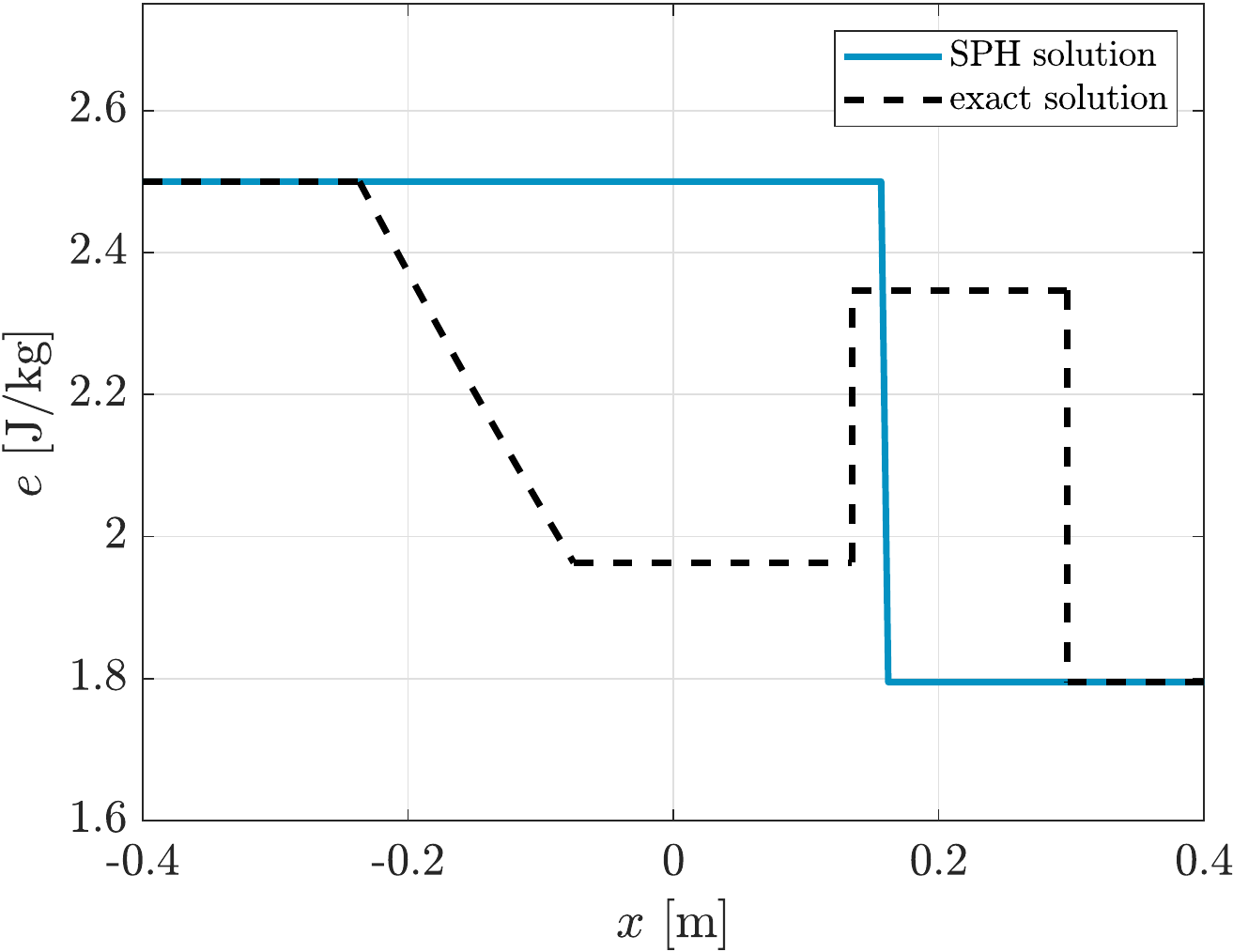}
        {\scriptsize (d) \emph{Energy profile}}
    \end{minipage}
    \caption{Profiles for the shock tube example presented in section~\ref{sec:sod_shock_tube}.}\label{fig:shocktube_profiles_1}
\end{figure}

The shock front is located around $ x = 0.3 $. The rarefaction wave is between $ x = -0.3 $ and $ x = 0 $.
The contact discontinuity is located between $ x = 0.1 $ and $ x = 0.2 $.

\subsection{Shear-driven cavity problem}\label{sec:shear_cavity}
Here, we consider the two-dimensional shear-driven cavity problem. This is the flow of a fluid in a squared section tube where the top side moves at a constant velocity $ v_{\mathrm{top}} $, while the other three sides are fixed. After a certain amount of time, the flow reaches a steady state, with the formation of a recirculation pattern.

For the numerical simulation in this section, the length of the side of the square domain is $ \ell = 10^{-3} $ m, the viscosity is $ \nu = 10^{-6} $ m$^{2}$/s, and the density is $ \rho = 10^{3} $ kg/m$^3 $. The top side of the square moves at a velocity of $ v_{\mathrm{top}} = 10^{-3} $ m/s.
The initial particle distribution is shown in Figure~\ref{fig:shear_cavity_time_0s}. A total of 1600 real particles (the filled circles~\tikz{\path[draw=black,fill=Sky] (0,0) circle (0.75mm);}) are evenly distributed in the square domain, while 320 virtual particles (the gray squares~\tikz{\path[draw=gray3,fill=gray1] (0,0) rectangle (1.5mm,1.5mm);}) are used on the boundaries. 

\begin{figure}[htbp]
  \centering
  \includegraphics[width=0.55\textwidth]{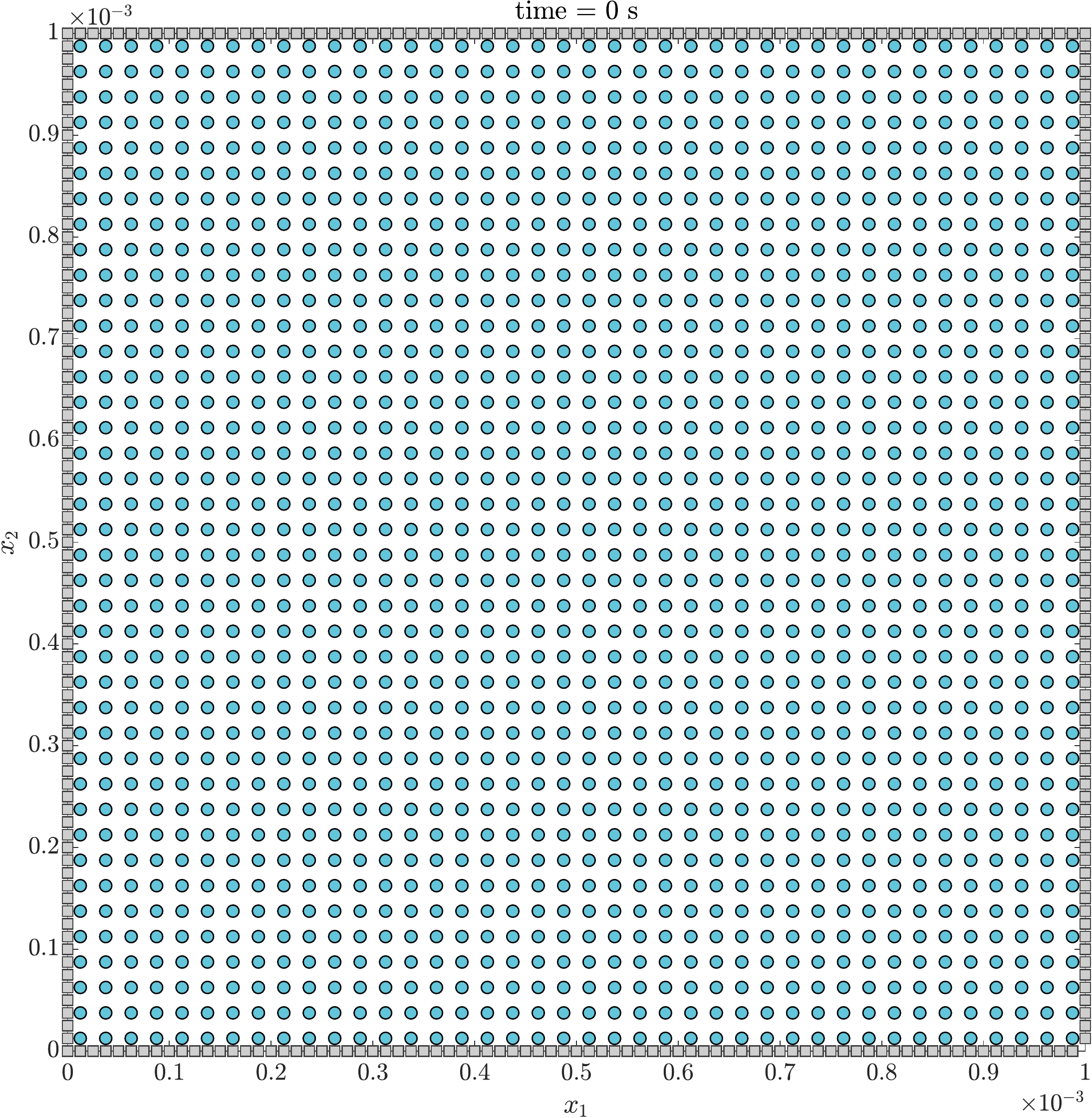}
  \caption{Initial particle distribution for the shear-driven cavity problem presented in section~\ref{sec:shear_cavity}.}\label{fig:shear_cavity_time_0s}
  \end{figure}

For the time evolution, we employ a constant time step $ \Delta t = 5 \times 10^{-5} $, and run the simulation for a total of 10\,000 time steps. Figure~\ref{fig:shear_cavity_steady_state} shows the particle and the velocity distribution when the steady state has been reached. The recirculation pattern can be clearly observed in panel (b) of Figure~\ref{fig:shear_cavity_steady_state}.
The results of this section can be reproduced by running the script \verb+Driver_SPHM+ with the parameter \verb+sph.example = 2+. Then, to generate Figure~\ref{fig:shear_cavity_steady_state}, run the script \verb+Driver_shear_cavity_steady_state+ with \verb@plt.velocity@ set to 0 or 1 to reproduce panel (a) or (b), respectively. 

\begin{figure}[htbp]
    \centering
    \begin{minipage}{0.48\textwidth}
        \centering
        \includegraphics[width=\textwidth]{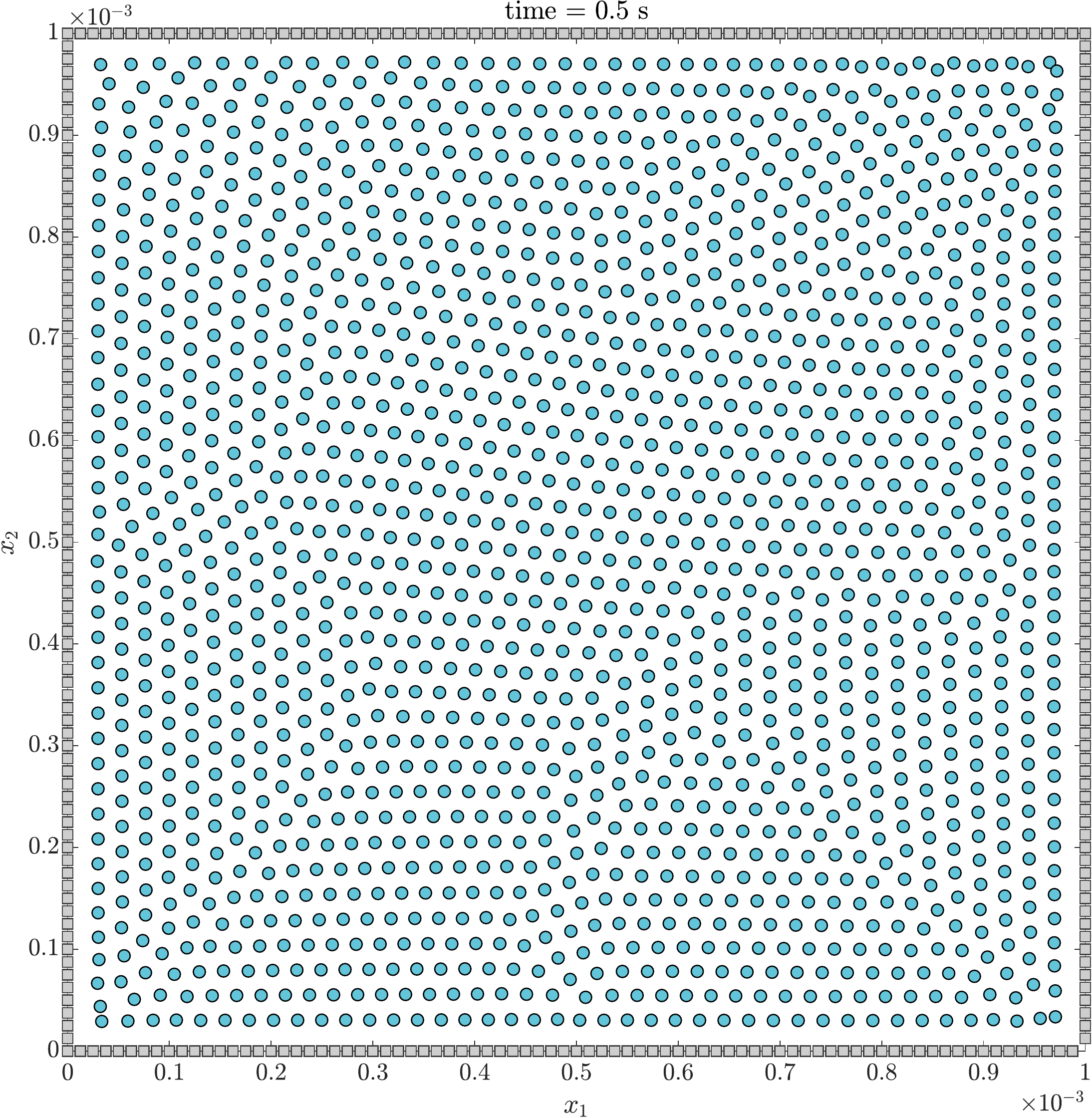}
        {\scriptsize (a) \emph{Positions}}
    \end{minipage}\hfill
    \begin{minipage}{0.48\textwidth}
        \centering
        \includegraphics[width=\textwidth]{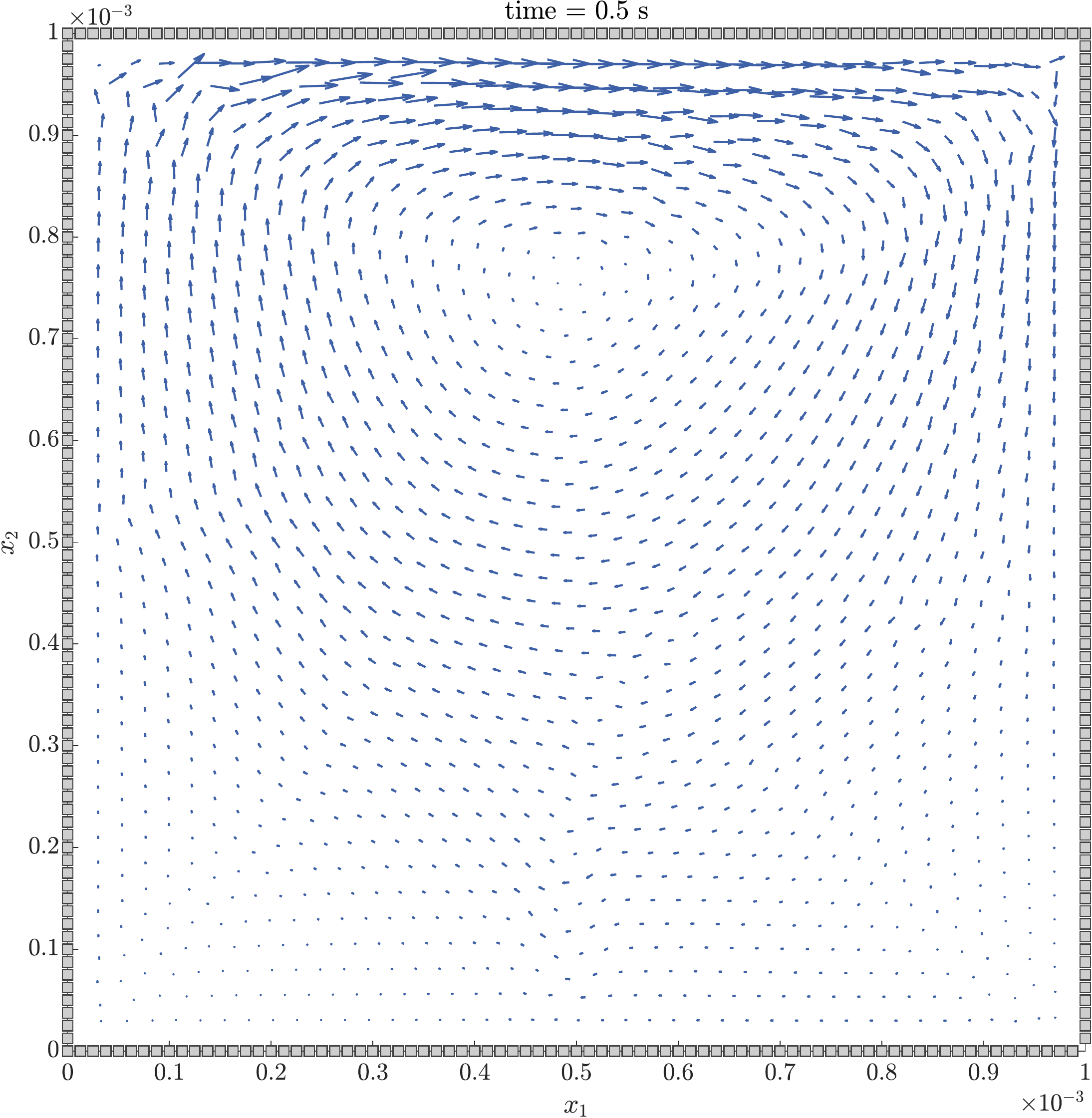}
        {\scriptsize (b) \emph{Velocities}}
    \end{minipage}
    \caption{Steady state particle and velocity distribution for the shear-driven cavity problem presented in section~\ref{sec:shear_cavity}.}\label{fig:shear_cavity_steady_state}
\end{figure}

\section{Floating objects}\label{sec:floating_obj}
The motion of a rigid body interacting with a fluid is determined by specifying the motion of the center of mass of the object and the rotation about the center of mass \protect{\cite{Monaghan:2005,Gesteira:2010}}. This is illustrated in Figure~\ref{fig:floating_object}, where $ \bX $ and $\mathbf{V} $ indicate the position and the velocity of the center of mass, respectively.

\begin{figure}[htbp]
\centering
\includegraphics[width=0.65\textwidth]{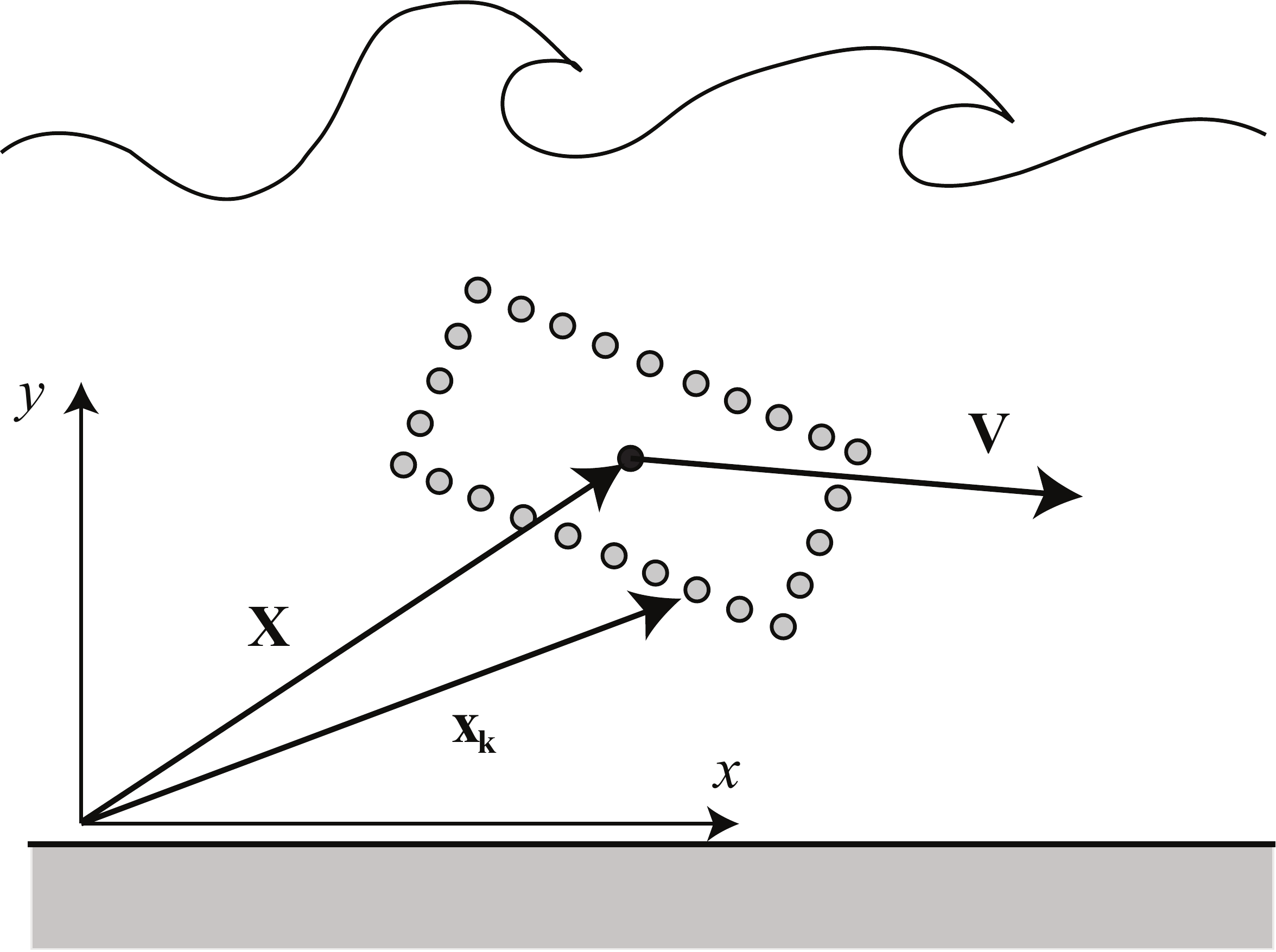} 
\caption{Floating object in a two-dimensional flow.}
\label{fig:floating_object}
\end{figure}

The floating object is represented by a rectangular rigid body, which is discretized by a set of boundary particles that are equally spaced around its boundary \protect{\cite{Monaghan:2003}}.

\subsection{Equations of motion}
The equations of motion for a floating object should describe the time evolution of the center of mass, the angular velocity, and the position of the particles on the boundary of the rigid body.

\paragraph{Center of mass.} Let $ B_{\mathrm{rb}} $ denote the set of the boundary particles of the rigid body. The center of mass of the rigid body is evolved according to the vector equation
\begin{equation}\label{eq:eom_mass}
   M \,\dfrac{\rmd\vel}{\dt}=\sum_{k \in B_{\mathrm{rb}}} m_{k}\,\fk,
\end{equation}
where $M$ is the mass of the rigid body, $m_{k}$ is the mass of the boundary particles of the rigid body and $\fk$ are the forces acting on the boundary particles of the rigid body due to the surrounding fluid. For the formulation of $\fk$, see section~\ref{sec:boundary_force}.

Figures~\ref{fig:floating_object2} and~\ref{fig:floating_object3} sketch the concept of boundary forces calculation for a floating object by using the normal boundary force approach. Note that~\eqref{eq:eom_mass} takes care of the three translational degrees of freedom of the rigid object.

\begin{figure}[htbp]
\centering
\includegraphics[width=0.55\textwidth]{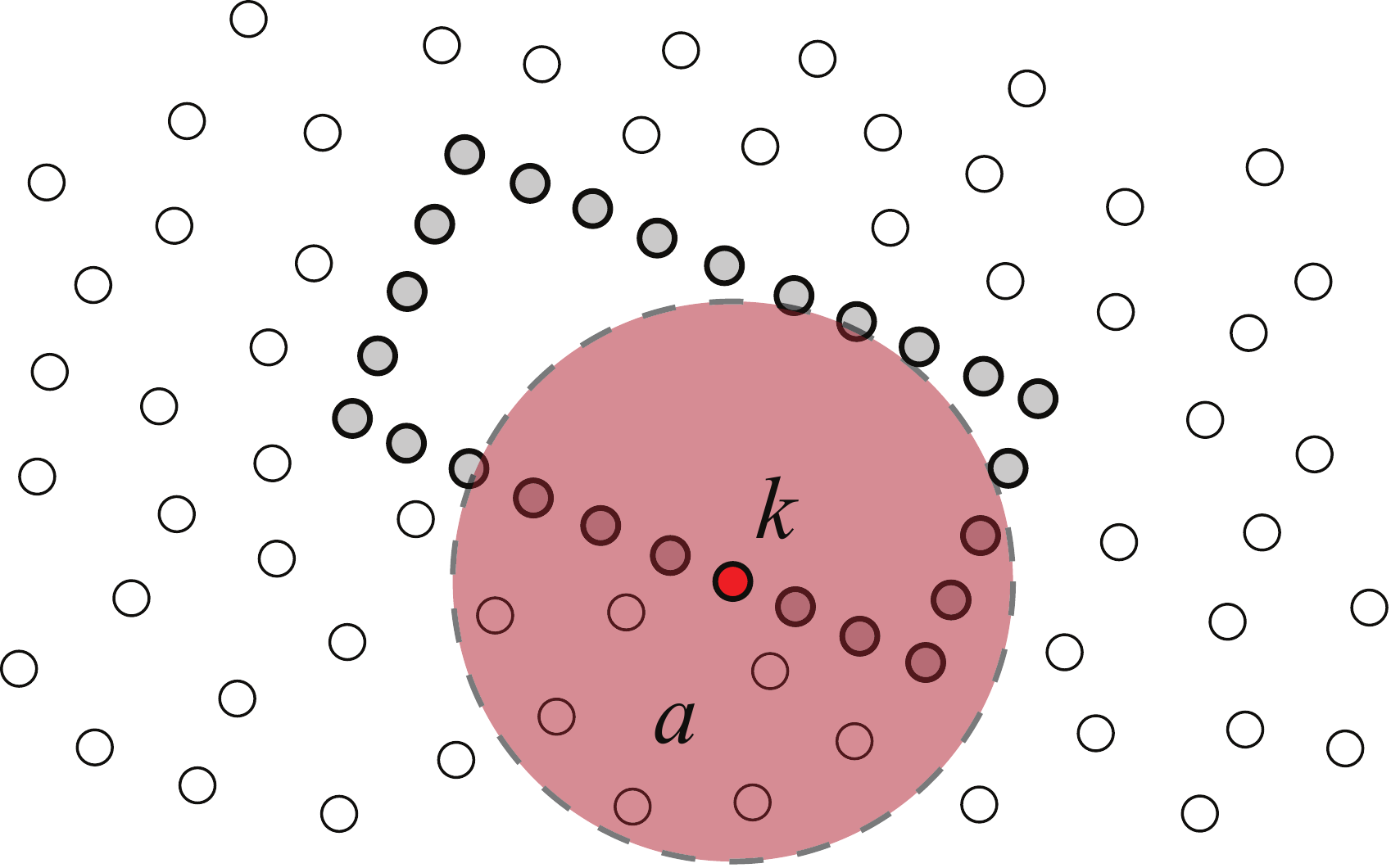} 
\caption{Calculation of the force on each boundary particle of the rigid body.}
\label{fig:floating_object2}
\end{figure}

\begin{figure}[htbp]
\centering
\includegraphics[width=0.45\textwidth]{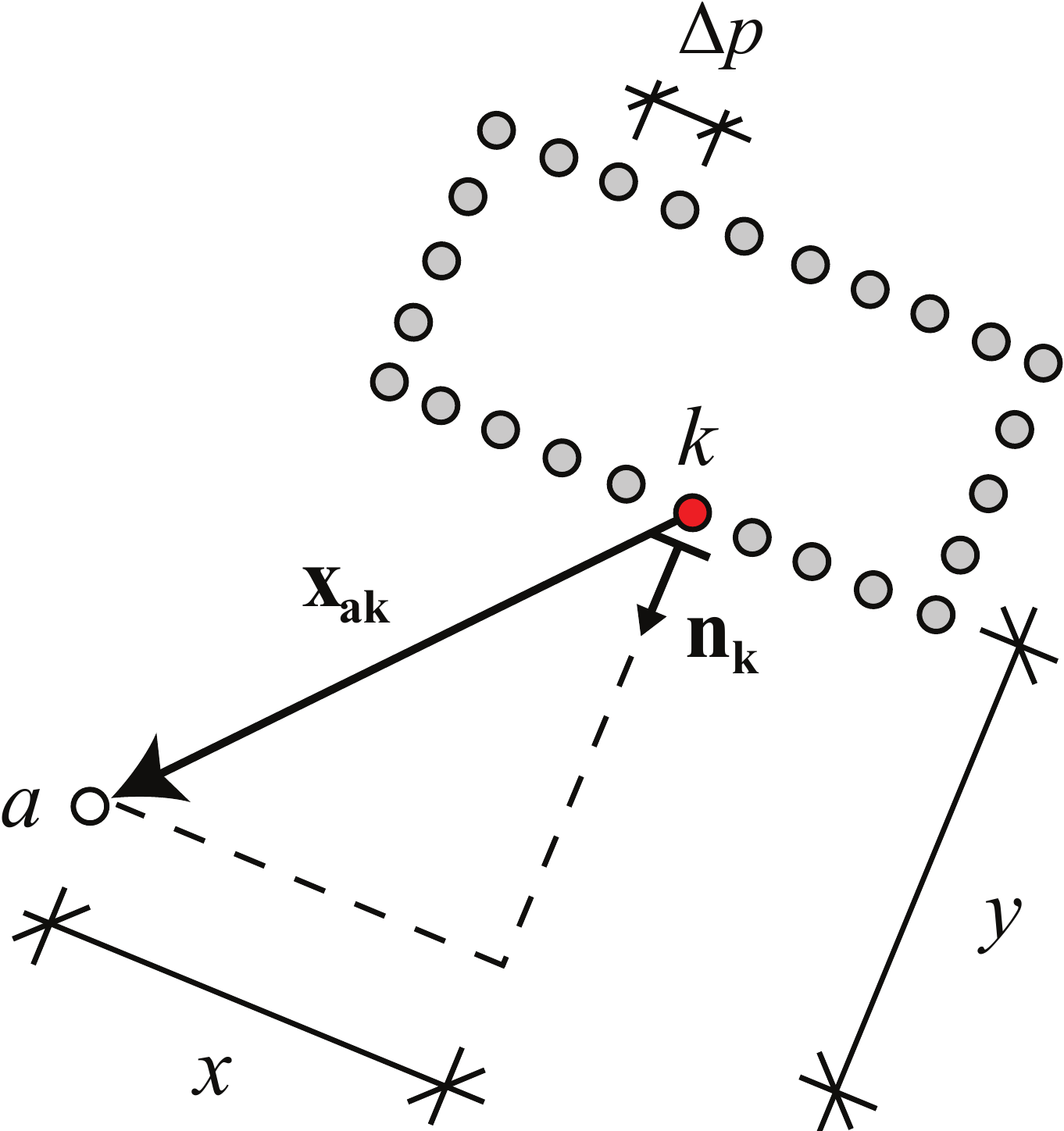} 
\caption{Calculation of the boundary force: force normal to the boundary.}
\label{fig:floating_object3}
\end{figure}

\paragraph{Angular velocity.}
In the case of a two-dimensional motion, the equation for the evolution of the angular velocity $ \bOmega $ is
\begin{equation}\label{eq:ang_vel}
   I\,\dfrac{\rmd\bOmega}{\dt}=\sum_{k \in B_{\mathrm{rb}}}m_{k}(\xk-\bX)\times \fk,
\end{equation}
where $I$ is the moment of inertia and $\xk$ denotes the position of the $k$-th boundary particle. This equation takes care of the rotational degree of freedom of the rigid object.
The term $ I \bOmega $ represents the \emph{angular momentum} (also called moment of momentum).

\paragraph{Boundary particles.} The boundary particles are moved according to
\begin{equation}\label{eq:boundary_particles}
   \dfrac{\rmd\xk}{\dt}=\vel+\bOmega\times(\xk-\bX).
\end{equation}
Equations~\eqref{eq:eom_mass}, \eqref{eq:ang_vel} and~\eqref{eq:boundary_particles} are integrated in time to predict the values of $\vel$, $\bOmega$ and the position $ \xk $ of the boundary particles for the next time step.

The SPH approximation of the equations of motion in presence of floating rigid objects becomes
\begin{equation*}
\dfrac{\dv_{i}^{a}}{\dt}=-\sum_{b=1}^{N}m_{b}\left( \dfrac{p_{a}}{\rho^{2}_{a}}+\dfrac{p_{b}}{\rho^{2}_{b}}\right) \dfrac{\partial W_{ab}}{\partial x_{i}^{a}} + 2 \sum_{b=1}^{N} m_{b} \left( \dfrac{\mu_{a} D'^{a}_{ij}}{\rho^{2}_{a}}+\dfrac{\mu_{b} D'^b_{ij}}{\rho^{2}_{b}}\right) \dfrac{\partial W_{ab}}{\partial x_{j}^{a}} + f_{i}^{a},
\end{equation*}
which is exactly equation~\eqref{eq:SPH_approx_eom_plus_boundary_force} that we get in section~\ref{sec:normal_boundary_force} when discussing \emph{fixed} boundaries. So the boundary force formulations that we discussed in the previous section are well suited for this kind of application.


\bibliographystyle{alpha_init}

\bibliography{SPHM_2022_biblio}

\end{document}